\documentclass[12pt]{iopart}
\usepackage{iopams} 
\usepackage{graphicx,epsf}

\def\n{{\tilde n}}
\def\f{\tilde f}
\def\Loop{{\rm loop}}

\def\Pflat{P^{\flat}}

\def\NeqFour{{{\cal N}=4}}
\def\Neqfour{{{\cal N}=4}}

\def\NeqEight{{{\cal N}=8}}

\def\tree{{\rm tree}}
\def\be{\begin{equation}}
\def\ee{\end{equation}}
\def\bea{\begin{eqnarray}}
\def\eea{\end{eqnarray}}
\def\nn{\nonumber}
\def\scut{{s\hbox{-}\rm cut}}
\def\tcut{{t\hbox{-}\rm cut}}

\def\spa#1.#2{\left\langle#1\,#2\right\rangle}
\def\spb#1.#2{\left[#1\,#2\right]}
\def\eps{\epsilon}

\def\MHVbar{$\overline{\hbox{MHV}}$}

\def\oneloop{{\rm 1\hbox{-}loop}}

\def\text#1{{\rm #1}}
\def\eqa{\begin{eqnarray}}
\def\eqae{\end{eqnarray}}

\def\fig#1{figure~{\ref{#1}}}

\def\eqn#1{eq.~(\ref{#1})}
\def\eqns#1#2{eqs.~({\ref{#1}}) and~({\ref{#2}})}

\def\sect#1{section~{\ref{#1}}}

\newbox\charbox
\newbox\slabox
\def\s#1{{      
        \setbox\charbox=\hbox{$#1$}
        \setbox\slabox=\hbox{$/$}
        \dimen\charbox=\ht\slabox
        \advance\dimen\charbox by -\dp\slabox
        \advance\dimen\charbox by -\ht\charbox
        \advance\dimen\charbox by \dp\charbox
        \divide\dimen\charbox by 2
        \raise-\dimen\charbox\hbox to \wd\charbox{\hss/\hss}
        \llap{$#1$} }}

\begin{document}

\title{Basics of Generalized Unitarity}

\author{Zvi Bern and Yu-tin Huang }

\address{
${}$Department of Physics, UCLA,
             Los Angeles, CA 94305-4060, USA
}

\ead{bern@physics.ucla.edu, yhuang@physics.ucla.edu}

\begin{abstract}
We review generalized unitarity as a means for obtaining loop
amplitudes from on-shell tree amplitudes.  The method is generally
applicable to both supersymmetric and non-supersymmetric amplitudes,
including non-planar contributions.  Here we focus mainly on
$\NeqFour$ Yang-Mills theory, in the context of on-shell superspaces.
Given the need for regularization at loop level,
we also review a six-dimensional helicity-based superspace formalism 
and its application to dimensional and
massive regularizations. An important feature of the
unitarity method is that it offers a means for carrying over any
identified tree-level property of on-shell amplitudes to loop level,
though sometimes in a modified form. We illustrate this with 
examples of dual conformal symmetry and a recently discovered duality
between color and kinematics.

This article is an invited review for a special issue of Journal of
Physics A devoted to ``Scattering Amplitudes in Gauge Theories''.
\end{abstract}

\maketitle


\section{Introduction}

The unitarity method~\cite{UnitarityMethod, DDimUnitarity,
  TwoLoopAllPlus, Z4Partons, GeneralizedUnitarity, TwoLoopSplit,
  BCFGeneralized, FiveLoop, OneLoopMethods,CompactThree, Neq44np,
  SixD} is at present the most powerful general means for obtaining
loop-level scattering amplitudes.  It applies to any massless
non-supersymmetric or supersymmetric theory, including their
non-planar contributions.  Here we will summarize basic ideas for
applying generalized unitarity as a means for obtaining loop
amplitudes from on-shell tree amplitudes.  We will focus on $\NeqFour$
super-Yang-Mills theory, in particular explaining how sums over
intermediate states are conveniently evaluated using on-shell
superspaces.  Up-to-date discussions of applications and further
details of generalized unitarity may be found in the chapters of this
review by Britto~\cite{BrittoReview}, by Carrasco and
Johansson~\cite{HenrikJJReview} and by Ita~\cite{ItaReview}.

The unitarity method has been applied to a wide variety of problems in
phenomenology and more theoretical aspects of quantum field theory.
This includes state-of-the-art predictions of LHC physics (for
example, see refs.~\cite{OneLoopMethods,QCDApplications,W4j}),
including the first next-to-leading-order QCD calculation with five
final state objects (including jets)~\cite{W4j}.  On the more
theoretical side, it has been used to open a new venue~\cite{BDS} for
studying Maldacena's AdS/CFT correspondence~\cite{Maldacena} and
perhaps eventually leading to a solution of planar $\NeqFour$
super-Yang-Mills theory. It has also been used to unravel a dual
conformal invariance in planar loop amplitudes of $\NeqFour$
super-Yang-Mills theory~\cite{DualConformal,KorchemskyOneLoop, BCDKS,
  RecentOnShellSuperSpace,LoopDualConformal, DualConformalTree,SixD,
  OConnelDualConf,DennenDual, DrummondReview,HennReview}.
Applications to supergravity
theories~\cite{GravityThree,CompactThree,GravityFour} have led to
results that have rejuvenated hope for the absence of ultraviolet
divergences for $\mathcal{N}=8$ supergravity. (For recent discussions
of this issue see refs.~\cite{GravityUVReview, SugraDiv,
  FreedmanReview}.)  More recently, it has also played a key role in
the conjecture of an all-loop-order duality between color and
kinematics and the associated double-copy structure of gravity
amplitudes in terms of gauge theory ones~\cite{BCJ,BCJLoop}.  Very
recently it has also played a role in a new means for recursively
finding multiloop multileg integrands in planar $\NeqFour$
super-Yang-Mills theory~\cite{PlanarIntegrand,NimaNew}.

The key feature of the unitarity method is that it constructs loop
amplitudes directly from on-shell tree amplitudes.  This makes it
possible to carry over any newly identified property or symmetry of
tree-level amplitudes to loop level.  This may be contrasted with
Feynman diagrammatic methods, whose diagrams are inherently gauge
dependent and off shell in intermediate states, making it difficult to
exploit any simplifying properties of the on-shell tree amplitudes.

Unitarity has a long history in quantum field theory since its
inception.  For a discussion of applications of unitarity during the
1960's, see ref.~\cite{ELOP}.  However, a variety of difficulties
prevented its widespread use as a means of constructing amplitudes, especially 
after the rise of gauge theories in the 1970's.  These difficulties
included the need for subtractions (meaning undetermined pieces), the
inability to construct amplitudes beyond four points and difficulties
with massless particles.  These basic difficulties were overcome with
the advent of the modern unitarity method~\cite{UnitarityMethod}. 

Over the years there have been a number of important refinements to
the unitarity method.  Generalized unitarity~\cite{ELOP} (where
multiple internal lines are placed on shell, subdividing a loop
amplitude into more than two pieces) was first applied in
ref.~\cite{Z4Partons} as a means for simplifying loop calculations. An
important more recent development is the use of complex
momenta~\cite{ComplexMomenta} by Britto, Cachazo and
Feng~\cite{BCFGeneralized}, leading to the realization that at one-loop in
four dimensions, quadruple cuts directly determine the coefficients of
all box integrals by freezing the loop integration.  Powerful new
methods for dealing with triangle and bubble integrals at one-loop, as
well as rational terms have also been developed~\cite{OneLoopMethods,
Bootstrap}. These are described in other chapters of this
review~\cite{BrittoReview,ItaReview}.  At higher loops, efficient
means of constructing the integrands of amplitudes, including
non-planar contributions, have also been
devised~\cite{FiveLoop,CompactThree,Neq44np,PlanarIntegrand}, as
discussed in some detail in another chapter of this 
review~\cite{HenrikJJReview}. 

Although the unitarity method applies just as well to supersymmetric
and non-supersymmetric theories, in general it is simpler to deal with
the supersymmetric case because of the simpler analytic structure of
such amplitudes.  In particular, it turns out that the better power
counting of supersymmetric theories allows all terms in one-loop
scattering amplitudes to be readily captured by cuts composed of
four-dimensional massless tree amplitudes.  Indeed, the first
applications of the unitarity method were for one-loop supersymmetric
amplitudes with arbitrary numbers of external legs, leading to
remarkably compact results~\cite{UnitarityMethod}.  For QCD or even
supersymmetric theories at higher loops, the situation is more
complex.  The source of this additional complexity is that
contributions that formally vanish in four dimensions may actually
contribute because they interfere with divergences. At one-loop
on-shell recursion relations can be used to construct the rational
terms, entirely in four dimensions~\cite{Bootstrap}. More generally,
by computing in $D=4 -2\eps$ dimensions all rational terms are
automatically captured~\cite{DDimUnitarity}.  A convenient way to
implement this, is by using six-dimensional
helicity~\cite{CheungOConnell,Boels,DHS}. This offers many of the
advantages of four-dimensional helicity, but meshes well with
regularization of the divergences, in a manner consistent with
unitarity~\cite{SixD}.  Besides regularization issues, there are a
number of interesting questions in higher-dimensional theories
accessible via six-dimensional techniques.  For example, it has
recently been used to confirm~\cite{SixD} that four-dimensional cuts
do capture all terms in the four-loop four-point amplitudes of
$\NeqFour$ super-Yang-Mills theory~\cite{Neq44np} and $\NeqEight$
supergravity~\cite{GravityFour}, which are relevant for higher-dimensional
studies of ultraviolet divergences in these theories.  Another
interesting application would be to confirm or refute the very
interesting recent proposal that maximally supersymmetric Yang-Mills
theory may be ultraviolet finite in five dimensions~\cite{Douglas}.

The present chapter of this review is organized as follows.  After
explaining the basics of the unitarity method, we illustrate this
method for the one-loop four-point amplitude of $\NeqFour$
super-Yang-Mills.  We start by working out the amplitude in
components.  Then we turn to more powerful superspace formalisms,
first in four dimensions and then in six dimensions. We then present
two nontrivial examples of how the unitarity method allows us to carry
over on-shell tree-level properties to loop level.  In the first of
these examples we show that dual conformal covariance of tree
amplitudes implies that loop-level integrands must transform nicely as
well.  In the second we will describe a duality between color and
kinematics, showing the transition from trees to loops.  In general,
tree properties carry over straightforwardly to generalized cuts that
decompose a loop amplitude into a product of tree amplitudes. However,
as the second example illustrates it can be nontrivial to demonstrate 
that a given
property holds for the complete loop amplitude.

\section{Basics of the unitarity method}
\label{BasicUnitaritySection}

\begin{figure}[th]
\centerline{\epsfxsize 4 truein \epsfbox{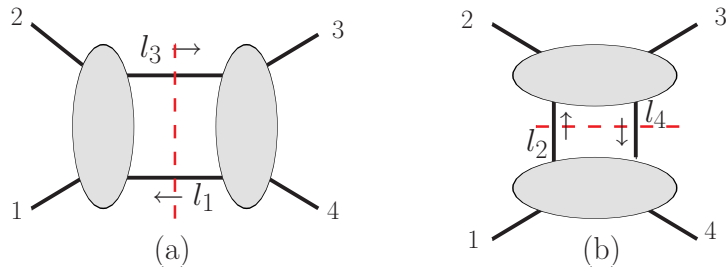}}
\caption[a]{\small The $s$ and $t$ channel 
two-particle cuts of the one-loop four-point 
amplitude.
}
\label{TwoParticleCutFigure}
\end{figure}

\subsection{Overview}

The simplest unitarity cuts to consider are the two-particle cuts. 
These are obtained by putting two intermediate lines on shell, as
illustrated in \fig{TwoParticleCutFigure} for a four-point one-loop 
amplitude. For example, the $s$ channel cut in \fig{TwoParticleCutFigure}(a)
is given by
\begin{equation}
C_s = \sum_{\rm states} A^\tree(-l_1, 1,2, l_3) \,
           A^\tree(-l_3, 3,4, l_1) \,,
\label{GeneralizedCutsChannel}
\end{equation}
where the sum runs over all physical states in the theory.
The cuts are evaluated using momenta that place all intermediate cut
momenta on shell, $l_i^2 = m_i^2$, where the $m_i$ are masses. 
Here we will take the theory to be massless.
Cuts are usually taken as including phase-space integrals, but for our
purposes it is simpler to define them as not including the phase-space
integration. 

An especially useful class of cuts are those that decompose a loop amplitude
into a sum over $m$ tree amplitudes of form,
\begin{equation}
C = \sum_{\rm states} A^\tree_{(1)} A^\tree_{(2)} A^\tree_{(3)} \cdots 
A^\tree_{(m)} \,,
\label{GeneralizedCut}
\end{equation}
where the sum runs over all physical states that can cross the cuts.
In $\NeqFour$ super-Yang-Mills theory, it is especially useful to
consider the maximal cuts~\cite{FiveLoop}, (also referred to as
``leading singularities''~\cite{FreddyMaximal}), where the maximum
number of propagator lines are placed on shell. 
Another useful class are single cuts where only a single internal line
is placed on shell~\cite{SingleCut}; these have played an important
role in the construction of planar integrands of $\NeqFour$
super-Yang-Mills theory via on-shell recursion~\cite{PlanarIntegrand}.

In general, the complete amplitude is determined from a ``spanning
set'' of cuts.  Such sets are found by considering all potential
independent contributions to the integrand that can enter an
amplitude (and which do not integrate to zero), based on power
counting or other constraints.  One simply needs to ensure that all
terms are non-vanishing in at least one cut that can then be used to
determine its coefficient.  In the $\NeqFour$ case one can often
construct an ansatz for the entire amplitude using various conjectured
properties.  Once one has an ansatz, by confirming it over the
spanning set, either numerically or analytically, we have a proof of
the correctness of the ansatz.

\begin{figure}[th]
\begin{center}
\centerline{\epsfxsize 5 truein \epsfbox{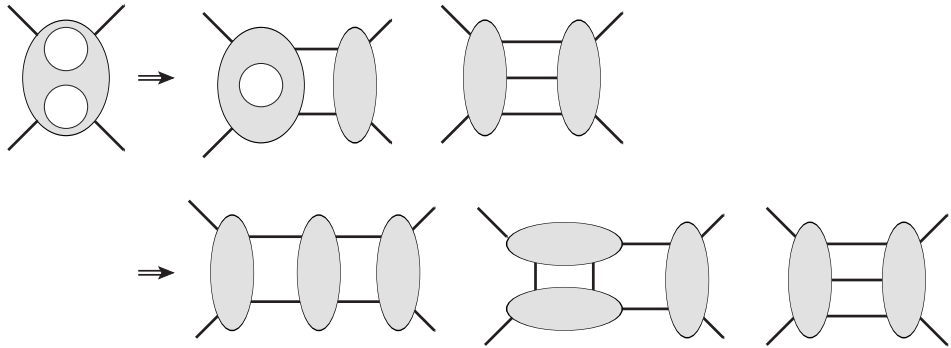}}
\caption[a]{\small The procedure for obtaining a
  spanning set of cuts for a planar two-loop four-point
   amplitude.  Only topologically distinct diagrams are
  shown.  The exposed intermediate lines are all placed on shell.  }
\label{TwoLoopFourPtFigure}
\end{center}
\end{figure}

One simple spanning set is obtained from the set of standard unitarity
cuts, where a given amplitude is split into two lower-loop amplitudes,
each with four or more external legs.  At $L$ loops, this is given by all cuts
starting from the two-particle cut to the $(L+1)$-particle cut in all
channels.   We can convert this to a
spanning set involving only tree amplitudes by iterating this process
until no loops remain. To illustrate this procedure, consider the
two-loop four-point amplitude displayed in
\fig{TwoLoopFourPtFigure}.  In the first stage this is decomposed 
using two- and three-particle cuts.  Then in the second stage, the one-loop
amplitude appearing in the two-particle cut is further decomposed into
its two-particle cuts.  In this figure we display only the
topologically distinct diagrams; the complete decomposition is
determined by considering all cuts with distinct labels.  If the
amplitudes are color ordered then we need to maintain a fixed
ordering of legs, depending on which planar or non-planar contribution
is under consideration.
On the other hand, if they are dressed with color then 
the distinct permutations of legs enter each cut automatically.

Another spanning set that is especially useful in $\NeqFour$
super-Yang-Mills theory is
obtained~\cite{FiveLoop,CompactThree,Neq44np} by starting from
``maximal cuts'', where the maximum numbers of internal propagators
are placed on shell.  These maximal cuts decompose the amplitudes into
products of three-point tree amplitudes, summed over the states
crossing the cuts. To construct a spanning set we systematically
release cut conditions one by one, first considering cases with one
internal line off shell, then two internal lines off shell and so
forth.  Each time a cut condition is released, potential contact terms
which would not be visible at earlier steps are captured.  The process
terminates when the only remaining potential contact terms exceed
power counting requirements of the theory (or integrate to zero in
dimensional regularization).  For the case of maximal
supersymmetry---especially in the planar case---there are a large
variety of additional tricks and techniques for obtaining
contributions efficiently~\cite{FiveLoop, FreddyMaximal, Neq44np,
  PlanarIntegrand, NimaNew}. 
Many of these are discussed further by Carrasco and Johansson in another
chapter of this review~\cite{HenrikJJReview}.

\begin{figure}[th]
\centerline{\epsfxsize 5 truein \epsfbox{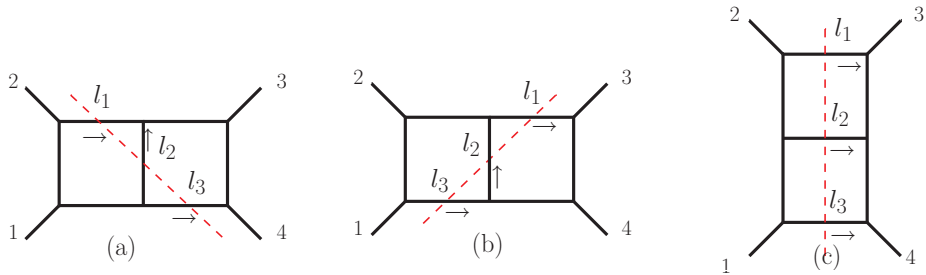}}
\caption[a]{\small The double-box three-particle cuts. The
 cut contributions (a) and (b) are two distinct cut contributions
 of the same horizontal double-box integral.  The vertical double box 
 has only a single contribution to the three-particle cut.}
\label{DoubleBoxCutsFigure}
\end{figure}

\begin{figure}[th]
\centerline{\epsfxsize 6.5 truein \epsfbox{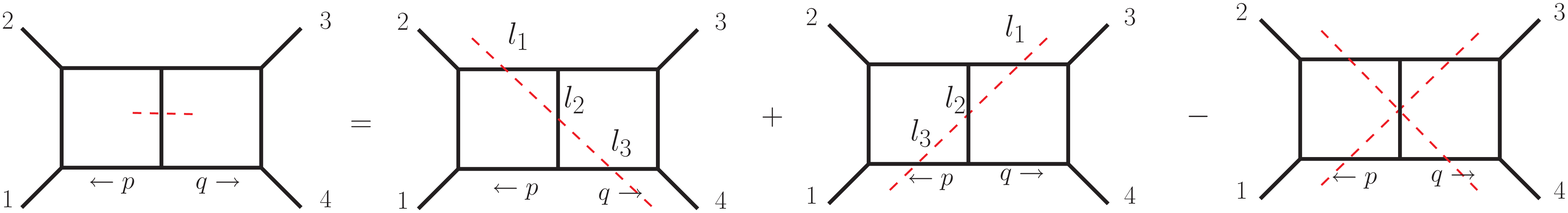}}
\caption[a]{\small An example showing how cuts are combined
to obtain contributions with fewer cut conditions. In this
equation the momentum labels of the different cut contributions need to 
be aligned, before being combined.  On the right hand side, the
only remaining cut conditions 
  }
\label{DoubleBoxCombineFigure}
\end{figure}

Given a spanning set of unitarity cuts, the task is to then find an
expression for the integrand of the amplitude with the correct cuts in
all channels.  This can be done either in a forward or reverse
direction.  In the forward direction the different cuts are merged
into integrands with no cut conditions using a merging procedure
described in ref.~\cite{TwoLoopSplit}. In the reverse direction we
first construct an ansatz for the amplitude containing unknown
parameters which are then determined by taking generalized cuts of the
ansatz and comparing to the cuts of the amplitude. The reverse direction
is usually preferred because we can expose desired properties,
simply by imposing them on the ansatz and then checking if its
unitarity cuts are correct.

\subsection{Cut merging procedure}

To illustrate the forward way of proceeding, consider a two-loop
four-point amplitude.  In particular, consider those contributions
which contain the propagators of the double-box integral displayed in
\fig{DoubleBoxCutsFigure},
\begin{equation}
\hskip -1.9 cm 
\int {d^{D}p\over (2\pi)^{D}} \;
 {d^{D}q\over (2\pi)^{D}} \; 
 {{N} \over p^2 \, (p - k_1)^2 \,(p - k_1 - k_2)^2 \,(p + q)^2 q^2 \,
        (q-k_4)^2 \, (q - k_3 - k_4)^2 } \,.
\end{equation}
The key task is to determine the
numerator factor $N$ from the unitarity cuts.  As illustrated in
\fig{DoubleBoxCutsFigure}, the three-particle cut in the $s_{12}$
channel will have three contributions.  The horizontal double box
gives two distinct contributions to the cut, while the vertical double
box gives only a single contribution.  When reassembling the
contribution from the cuts we need to account for this relative double
count in the horizontal box.  A procedure for doing so is by adding
together the two horizontal box contributions, but subtracting out the
overlap, to prevent a double count, as illustrated in
\fig{DoubleBoxCombineFigure}. 

 Prior to adding the two contributions
(a) and (b) in \fig{DoubleBoxCutsFigure} we must switch from cut
labels used in diagram (a) to the diagram momentum labels, $p$ and
$q$, so that uniform momentum labels are used when combining the
contributions.
Note that the contributions (a) and (b) each require a different change
of variables. For diagram (a) we require
\begin{equation}
l_1 = p - k_1 - k_2\,, \hskip 1 cm l_2 = -p-q\,, \hskip 1 cm
l_3 = q\,, 
\end{equation}
while for diagram (b) we require
\begin{equation}
l_1 = -q +k_3 +k_4\,, \hskip 1 cm l_2 = p+q\,. \hskip 1 cm
l_3 = -p \,,
\end{equation}
The third term on the right hand side of \fig{DoubleBoxCombineFigure}
is the ``overlap'' between the numerator contributions (a) and (b)
which must be subtracted out to prevent double counts.  The overlap
term is determined by taking contribution (a) and imposing the
additional cut conditions of diagram (b).  In carrying this out, it is
important that terms which vanish by the cut conditions are set to
zero.  This is best accomplished by expressing the numerator in terms
of inverse propagators, whenever possible.  A key consistency
condition is that the overlap terms need to be the same if we swap the
roles of (a) and (b).  (In explicit calculations it is helpful to
always check this consistency condition, since it will catch most
errors.)

This merging procedure generalizes to any set of cuts. In general, we
can write a gauge-theory amplitude as a sum over integrals,
\begin{equation} 
{\cal A}_n^{(L)}= g^{2L + n -2} \sum_{{\rm diags.}\ i} {1\over S_i} c_i I_i\,,
\end{equation}
where the $S_i$ are symmetry factors, the $c_i$ color factors, the
$I_i$ some loop integrals and the sum runs over all contributing integrals. 
The loop integrals are of the form,
\begin{equation}
I_i =  \int \prod_{j=1}^L \frac{d^D l_i}{(2\pi)^D}
   { n_i \over \prod_{j \in \alpha_i} p_j^2},
\end{equation}
where the $n_i$ are numerators.  The numerators can be local 
or non-local depending on the chosen organization.
The product in the denominator runs over all propagators in the
graph.  Each cut contribution is then assigned to a graph according to
the propagators it contains and the merging procedure is carried out
for each graph.  For each set of propagators specifying an integral
function, one must add together the contributions from each cut,
subtracting any overlap.  A detailed discussion of this procedure may
be found in ref.~\cite{TwoLoopSplit}. 

We note that the merging procedure is not limited to planar supersymmetric
theories and has been successfully applied to QCD~\cite{TwoLoopSplit}.
In special cases, such as $\NeqFour$ super-Yang-Mills theory at one
loop, the merging procedure is rather simple, because we can make use
of the fact that the amplitude is determined solely from box
integrals~\cite{UnitarityMethod}, whose coefficients may be determined
directly from quadruple cuts~\cite{BCFGeneralized}.

\subsection{Merging cuts via an ansatz}

In practical calculations, it is usually simpler to work in the
reverse direction, by first constructing an ansatz for the numerators
containing all potential terms that can appear.  The numerators can
depend on the momenta, polarizations and spinors and contain
parameters to be determined from the cuts. When using superamplitudes, 
the numerators will depend on anticommuting parameters as well.

To determine the unknown numerical coefficients we take generalized
cuts of the ansatz and systematically set these equal to the cuts of
the loop amplitude. Usually it is best to enforce a set of auxiliary
constraints, such as demanding an upper limit to the number of powers
of loop momenta that that can appear in the numerators of the
integrands.  Depending on the situation, one can, for example, impose
that each numerator is a polynomial in the loop and external momenta.
One can greatly reduce the number of cuts required to determine the
amplitude by feeding in additional information, such as dual conformal
symmetry for the planar $\NeqFour$ super-Yang-Mills
amplitudes~\cite{FiveLoop,NimaNew}.  This procedure is simplest for
the four-point amplitudes of $\NeqFour$ super-Yang-Mills theory, where
it turns out that the ratio between the loop integrand and the tree
amplitudes is a rational function solely of Lorentz invariant scalar
products, at least through four loops and very likely beyond
this~\cite{BRY,BCDKS, FiveLoop,Neq44np}.  For higher-point amplitudes,
similar ratios can contain either spinor products or Levi-Civita
tensors.  If it turns out that no solution to the cut conditions is
found, the ansatz then needs to be enlarged by removing auxiliary
constraints.

A given cut configuration will not completely fix the ansatz as the
numerator may be freely modified by adding terms that vanish on the
cut under study.  Consider, for example, a particular two-particle cut
with cut momenta labeled $l_1$ and $l_2$.  Expressions proportional to
$l_1^2=0$ and $l_2^2=0$ are not constrained by this particular cut.
Such terms are instead constrained by other cuts.  After information
from all cuts is included, the only remaining ambiguities are terms
that are free of cuts in {\it every} channel.  In the full amplitude
these ambiguities add up to zero, representing the freedom to
re-express the amplitude in different equivalent forms. This
freedom allows us to find representations with different desirable
properties, such as manifest symmetries or power
countings~\cite{CompactThree, BCJLoop, Neq44np}.

For multiloop calculations in $\NeqFour$ super-Yang-Mills theory, it
is usually helpful to organize the evaluation of the cuts following
the method of maximal cuts~\cite{FiveLoop}.  This allows us to focus
on a relatively small number of terms, as well as reducing the
complexity of each cut.  A related procedure, valid for maximally
supersymmetric amplitudes, makes use of
``leading-singularities''~\cite{FreddyMaximal}.  These leading
singularities, which include additional hidden singularities, likely
determine any maximally supersymmetric amplitude, once additional
symmetry information is incorporated~\cite{AHCKGravity,NimaNew}.

Similar constructions can also be done in supergravity theories.
However, it is generally simpler to construct loop amplitudes in these
theories by first evaluating corresponding super-Yang-Mills loop
amplitudes which can then be using directly in the evaluation of the
supergravity cuts~\cite{BDDPR,GravityThree,GravityFour}, using either
the tree-level KLT~\cite{KLT} relations or the double-copy BCJ
relations~\cite{BCJ}. In fact, it appears likely that one can read off
multiloop supergravity amplitudes directly from the corresponding
gauge-theory amplitudes after they are arranged in a form that
satisfies a color-kinematics duality~\cite{BCJLoop}.

\subsection{Regularization issues}

One important technical issue is that amplitudes are in general
divergent and need to be regularized.  For most purposes, the most
convenient regularization scheme is dimensional regularization.  For
planar $\NeqFour$ super-Yang-Mills theory only infrared divergences
are present, and in this case a massive Higgs regulator has proven to
be especially convenient~\cite{HiggsReg}. Interestingly, this form of
massive regularization is closely related to dimensional
regularization, and obtained by treating the extra dimensional components
of momenta as masses not to be integrated over~\cite{SixD}.

At one loop, all contributions to the amplitudes of massless
supersymmetric theories are determined completely by their
four-dimensional cuts~\cite{UnitarityMethod}.  Unfortunately, no such
theorem has been proven at higher loops.  There is, however,
substantial evidence that it holds for four-point amplitudes in this
theory through six
loops~\cite{BCDKS,GravityThree,FiveLoop,Neq44np,SixD}.  We do not
expect that it will continue for higher-point amplitudes.  Indeed, we
know that for two-loop six-point amplitudes, terms which vanish in
$D=4$ do contribute in dimensional regularization~\cite{TwoLoopSixPt}.
However, it may well turn out that all terms not determined by
four-dimensional massless cuts have a predictable universal structure,
at least near four dimensions.  Interestingly, experience indicates
that when using the massive regulator, we can safely drop all terms of
order $m$, where $m$ is the effective mass obtained from the 
extra-dimensional components~\cite{HiggsReg,PlanarIntegrand,KRV}.

Until theorems are proven determining when four-dimensional massless
cuts are sufficient to guarantee that all terms are captured, 
by evaluating cuts in higher dimensions we can ensure their reliability.  In
general, using $D$-dimensional
momenta~\cite{DDimUnitarity,TwoLoopAllPlus} makes calculations
significantly more clumsy, because powerful spinor and superspace
methods~\cite{SpinorHelicity} are no longer applicable.  However,
thanks to refs.~\cite{CheungOConnell,DHS}, we now have helicity and
superspace formalisms in six dimensions analogous to the well known
four-dimensional ones.  In many cases this is sufficient, allowing for
straightforward constructions of dimensionally or massively
regularized amplitudes~\cite{SixD}.  This then offers a convenient way
to construct regularized integrands that are guaranteed to be
consistent with unitarity.

\section{Sample Calculation: One-loop four-point amplitude in 
maximal super-Yang-Mills theory}
\label{ComponentsSampleSection}

For relatively simple amplitudes, the sum over states in a cut
(\ref{GeneralizedCut}) can be evaluated straightforwardly in
components~\cite{BDDPR}.
As we
discuss in subsequent sections, for more complex situations where the
bookkeeping becomes more difficult, the sums over states are best
handled~\cite{FreedmanGenerating, FreedmanUnitarity, SuperSum,
  SixD,SixDTwo,BSTReview} using on-shell superspaces~\cite{Nair, Boels, DHS}.

As a simple example where a component analysis is adequate, consider a
four-gluon amplitude of $\NeqFour$ super-Yang-Mills theory. This
amplitude was first computed in ref.~\cite{GSB}, as the low-energy
limit of a type~I string.  For the purposes of illustrating the cut
construction, we may view this amplitude as an ordinary gauge-theory
amplitude but with a particular matter content: one gluon, four
gluinos and six real scalars all in the adjoint representation.  In
supersymmetric theories amplitudes with identical helicity or with one
leg of opposite helicity (in an all outgoing convention) vanish to all
loop orders by a supersymmetry identity~\cite{SWI}.  (A discussion of
supersymmetry identities can be found in another chapter of this
review~\cite{FreedmanReview}.)
The first non-trivial case to consider is the
four-point case with two negative and two positive helicities,
$A_{4}^\Neqfour(1^-, 2^-, 3^+, 4^+)$.

In general, we can write the color-ordered amplitude as
\begin{equation}
\hskip -2 cm A_{4}^\Neqfour(1^-, 2^-, 3^+, 4^+) 
= - i A_4^\tree   \int {d^{4-2\eps}p\over (2\pi)^{4-2\eps}} \;
{ N \over p^2 (p - k_1)^2 (p - k_1 - k_2)^2 (p + k_4)^2}\,,
\end{equation}
where we have extracted an overall factor of the tree amplitude.  (The
amplitude is the coefficient of the color trace $N_c \tr(T^{a_1}
T^{a_2} T^{a_3} T^{a_4})$ for an SU($N_c$) gauge group.)  Our
task is to determine the numerator factor $N$ from the cuts.  We have
chosen to write the amplitude entirely as a box integral.  In this
representation, if any triangle or bubble contributions were to appear
they would enter as inverse propagators in the numerator canceling
propagators.

For simplicity, we will work this out using four-dimensional
momenta. As mentioned above, in general, this can lead to terms being
dropped, but supersymmetric amplitudes at one loop are known not to
contain additional terms~\cite{UnitarityMethod}.  In carrying out this
calculation we assume the relevant tree-level amplitudes are already
known, so we just quote their form~\cite{TreeReview}.  First consider
the $s$-channel cut depicted in \fig{TwoParticleCutFigure}(a). 
The necessary tree amplitudes are the four-gluon
amplitudes,
\begin{eqnarray}
&& A_4^\tree(-l_1^+, 1^-, 2^-, l_3^+) = i { \spa1.2^4 \over
\spa{-l_1}.1 \spa1.2 \spa2.{l_3} \spa{l_3}.{-\!l_1}}\,,\nn \\
&& A_4^\tree(-l_3^-, 3^+, 4^+, l_1^-) = i {
\spa{-l_1}.{l_3}^4 \over \spa{-l_3}.3 \spa3.4 \spa4.{l_1}
\spa{l_1}.{-\!l_3}} \,,
\label{TreeFourAmpls}
\end{eqnarray}
where $\spa{i}.{j}$ denote spinor inner products following the
notation of ref.~\cite{LanceReview}. 
All other combinations of
helicities of the intermediate lines cause at least one of the tree
amplitudes on either side of the cut to vanish.  (The
outgoing-particle helicity convention means that the helicity label
for each intermediate line flips when crossing the cut.)
Remarkably, for
this cut only the gluon loop contributes; for fermion or scalar loops
at least one of the two tree amplitudes vanish. 
Using the
tree amplitudes (\ref{TreeFourAmpls}), the cut in the $s$ channel
depicted in \fig{TwoParticleCutFigure}(a) is simply,
\begin{eqnarray}
C_s & = &
 {i \spa1.2^4 \over \spa{l_1}.1 \spa1.2 \spa2.{l_3} 
             \spa{l_3}.{l_1}} 
 \; {i\spa{l_1}.{l_3}^4
\over \spa{l_3}.3 \spa3.4 \spa4.{l_1} \spa{l_1}.{l_3}}
 \,, 
\label{SCutSusyA}
\end{eqnarray}
where we used the assignment that 
$|\! -p\rangle=-|p\rangle$ and $|\!-p]=|p]$ 
such that $|\! -p\rangle[-p|=-p$. To put this into a form more
reminiscent of integrals encountered in Feynman diagram calculations
we rationalize the denominators using, for example,
\begin{equation}
 {1\over \spa2.{l_3}} = -{\spb2.{l_3} \over (p - k_1)^2} \, ,
\label{Rationalize}
\end{equation}
where we set $l_1 = p$, $l_3 = p - k_1 - k_2$ and we used the on-shell
conditions $l_1^2=l_3^2=0$.  Performing these simplifications
yields
\begin{equation}
C_s =  i A_4^\tree  { {\tilde N} \over (p - k_1)^4 (p + k_4)^4}  \,,
\label{SCutSusyB}
\end{equation}
where we have extracted a overall factor of the tree amplitude from
the amplitude,
\begin{equation}
A_{4}^\tree(1^-, 2^-, 3^+, 4^+) =
 i {\spa1.2^4 \over \spa1.2 \spa2.3 \spa3.4 \spa4.1} \,.
\label{ggggmmpptree}
\end{equation}
The numerator $\tilde N$ is given by
\begin{eqnarray}
{\tilde N} & =& \spb{l_1}.1 \spa1.4 \spb4.{l_1} \spa{l_1}.{l_3}
             \spb{l_3}.3 \spa3.2 \spb2.{l_3} \spa{l_3}.{l_1} \nn \\
&=& \tr_+ [l_1 1 4 l_1 l_3 3 2 l_3] \nn \\
&=& - 4 \tr_+[4321]\,  l_1 \cdot k_1 \, l_1\cdot k_4
           = -st \, (p - k_1)^2 (p + k_4)^2 \,, 
\end{eqnarray}
where $\tr_+[\cdots] = {1\over 2} \tr[(1+\gamma_5) \cdots]$.
To simplify the expression we used,
\begin{equation}
l_1^2 = 0 \, , 
\hskip 1 cm \s l_1 \s l_3 = \s l_1 (\s k_3 + \s k_4) \, , 
\hskip 1 cm \s l_3 \s l_1 = -(\s k_1 + \s k_2) \s l_1 \, . 
\end{equation}
The $\gamma_5$ term in the trace drops out because 
$k_4$ is linearly dependent on the other external 
momenta and drops out when contracted into the
totally anti-symmetric Levi-Civita tensor.

Combining the equations and putting back the cut
propagators, we find that
\begin{equation}
A_{4;1}^\Neqfour(1^-, 2^-, 3^+, 4^+)\Bigr|_{\scut} 
= - s t \, A_4^\tree  \, I_4(s,t)
 \Bigr|_{\scut} \,,
\label{SchannelCut}
\end{equation}
has the correct $s$-channel cut, 
where $I_4(s,t)$ is the box integral,
\begin{equation}
I_4(s,t) = -i \int {d^{4-2\eps}p\over (2\pi)^{4-2\eps}} \;
{1\over p^2 (p - k_1)^2 (p - k_1 - k_2)^2 (p + k_4)^2} \,. 
\label{BoxIntgegral}
\end{equation}

Now consider the $t$ channel shown in \fig{TwoParticleCutFigure}(b). In
this case, all particles in the multiplet contribute to the cut.  We 
label the states according to their helicity $h$.
In this case the two
tree-level amplitudes on either side of the cuts are given by
\begin{eqnarray}
&& A_4^\tree(-l_2^{-h}, 2^-, 3^+, l_4^{h}) = i { 
\spa{-l_2}.{2}^{2 + 2h} \spa{l_4}.{2}^{2-2h} 
 \over
\spa{-l_2}.2 \spa2.3 \spa3.{l_4} \spa{l_4}.{-\!l_2}}\,, \nn \\
&& A_4^\tree(-l_4^{-h}, 4^+, 1^-, l_2^{h}) = i {
\spa{-l_4}.{1}^{2+2h}\spa{l_2}.{1}^{2-2h}
 \over \spa{-l_4}.4 \spa4.1 \spa1.{l_2}
\spa{l_2}.{-\!l_4}} \,.
\end{eqnarray}
In $\NeqFour$ super-Yang-Mills theory the helicity $h$ takes on 
values, 
$\{-1,-1/2,0,1/2,1\}$. 
These amplitudes can be obtained from the purely gluonic ones,
using supersymmetry Ward Identities~\cite{SWI,FreedmanReview}
or by extracting the components from superspace amplitudes, discussed
below.

The $t$-channel cut is given by multiplying the two amplitudes and summing 
over all intermediate states.  This gives us
\begin{equation}
C_t = - \rho {1 \over
    \spa{-l_2}.2 \spa2.3 \spa3.{l_4} \spa{l_4}.{-\!l_2}} \,
{1 \over \spa{-l_4}.4 \spa4.1 \spa1.{l_2} \spa{l_2}.{-\!l_4}} \,,
\end{equation}
where the factor $\rho$ contains the contributions from 
the gluon, four gluinos and six real scalars,
\begin{eqnarray}
\rho &=& \Bigl(\spa{l_2}.{2}^4 \spa{l_4}.{1}^4
    - 4 \spa{l_2}.{2}^{3} \spa{l_4}.{2} 
                           \spa{l_4}.{1}^3\spa{l_2}.{1}  \nn\\
  &&\null
    + 6 \spa{l_2}.{2}^2 \spa{l_4}.{2}^2 
                             \spa{l_4}.{1}^{2}\spa{l_2}.{1}^{2} 
    - 4\spa{l_2}.{2} \spa{l_4}.{2}^3 \spa{l_2}.{1}^{3}\spa{l_4}.{1} \nn\\
&&  \hskip 3 cm 
    + \spa{l_4}.{2}^{4}\spa{l_2}.{1}^4 \Bigr) \nn\\
 &=& \Bigl(\spa{l_2}.{2} \spa{l_4}.{1}
       - \spa{l_4}.{2}\spa{l_2}.{1}\Bigr)^4 \nn \\
 &=& \spa{1}.{2}^4 \spa{l_4}.{l_2}^4\,.
\end{eqnarray}
To obtain the final line we 
made use of a Schouten identity~\cite{TreeReview}.
Thus, the $t$ channel cut collapses to a 
relabeling of the $s$ channel cut (\ref{SchannelCut}).

Following similar algebra as for the $s$ channel cut
we obtain
\begin{equation}
A_{4}^\Neqfour(1^-, 2^-, 3^+, 4^+)\Bigr|_{\tcut} 
= {- s t }\, A_4^\tree  \, I_4(s,t)
 \Bigr|_{\tcut} \,, 
\label{TchannelCut}
\end{equation}
At this point, from \eqns{SchannelCut}{TchannelCut}
it is clear that the correct amplitude must be 
\begin{equation}
A_{4}^\Neqfour(1^-, 2^-, 3^+, 4^+)
= - s t \, A_4^\tree  \, I_4(s,t) \,, 
\label{FourGluonAmplitude}
\end{equation}
because it has the correct cuts in both the $s$ and $t$
channels.  In this case the merging procedure for the cuts
turns out to be trivial
because the numerator does not contain loop momentum.

More complex cases both at one and higher loops are discussed in other
chapters of this review~\cite{HenrikJJReview, BrittoReview,
ItaReview}.  Recent progress has also been made using cuts with only a
single cut propagator~\cite{SingleCut,PlanarIntegrand}.

\section{Sewing  $\Neqfour$ Super-Yang-Mills amplitudes}


For problems involving supersymmetric amplitudes, it is usually best
to use an on-shell superspace which organizes the amplitudes
according to physical helicity states.  This provides a convenient means for
dealing with all states of the theory simultaneously, and for carrying
out intermediate sums of states crossing cuts.  The superspace we
discuss here consists of unconstrained Grassmann 
variables which form a fundamental representation of the
SU$(\mathcal{N})$ R-symmetry group. The unconstrained nature of the
variables makes it simple to translate summations of on-shell states
needed in unitarity cuts into Grassmann integrations,
which take care of the bookkeeping.

\subsection{Tree-level superamplitudes}
\label{SuperTreeSubsection}

We begin with a brief setup of the formalism.  (See also the chapter
of this review from Brandhuber, Spence and Travaglini~\cite{BSTReview}.)
The on-shell superspace is constructed by introducing a
 set of fermionic variables $\eta^a$, with $a=1 \cdots \mathcal{N}$
transforming in the fundamental representation of SU$(\mathcal{N})$
R-symmetry. The bosonic spinor variables carry kinematic
information, while the fermionic variables carry information on the
helicity and R-symmetry representation of the external states.
$\Neqfour$ super-Yang-Mills has a simple structure 
because all states can be incorporated into a self CPT 
superfield,
\begin{equation}
\hskip -.5 cm 
\Phi(\eta)=g^++\eta^af^+_a+\frac{1}{2}\eta^a\eta^b\phi_{ab}+\frac{1}{3!}\epsilon_{abcd}\eta^a\eta^b\eta^cf^{d-}+\frac{1}{4!}\epsilon_{abcd}\eta^a\eta^b\eta^c\eta^dg^-\,.
\label{MHV}
\end{equation}   
We note that $\NeqEight$ supergravity is similar,
with a self CPT multiplet in four dimensions 
containing fields up to spin 2.

These fermionic variables were first introduced by
Ferber~\cite{Ferber1977qx} to extend twistors, which are
representations of four-dimensional conformal group, to
supertwistors. Nair~\cite{Nair} applied these variables to represent
MHV amplitudes of $\Neqfour$ super-Yang-Mills theory. These amplitudes take
the form
\begin{equation}
\mathcal{A}^{\rm MHV}_n(1,2,\cdots,n)=\frac{i}{\prod_{j=1}^n\langle j(j+1)\rangle}\,
\delta^{(8)}\Biggl(\sum_{j=1}^n\lambda^{\alpha}_j\eta_j^a\Biggr)\,,
\label{MHVSuperAmplitude}
\end{equation}   
where leg $n+1$ is to be identified with leg $1$, and 
\begin{equation}
\delta^{(8)}\Biggl(\sum_{j=1}^n\lambda^{\alpha}_j\eta_j^a\Biggr)
= \prod_{a=1}^4\sum_{i<j}^n\langle ij\rangle\eta_i^a\eta_j^a\,.
\label{DeltaSpinor}
\end{equation}
The component amplitudes are the coefficients in the $\eta$ expansion
of $\mathcal{A}_n$, with the external states identified according to
their organization within the superfield as in \eqn{MHV}.

An MHV superamplitude of  $\mathcal{N}=4$ super-Yang-Mills can 
also be written as
 \begin{equation}
 \mathcal{A}_n^{\rm MHV}=\frac{A^\tree_{n}(1^-,2^-,3^+,\cdots,n^+)}{\langle 12\rangle^4}\,
\delta^{(8)}\Biggl(\sum_{j=1}^n\lambda^{\alpha}_j\eta_j^a\Biggr)\,,
 \label{MHVsYM}
 \end{equation}
where $A_{n}^\tree(1^-,2^-,3^+,\cdots,n^+)$ is the tree-level MHV
pure-gluon amplitude. We note that the MHV
superamplitudes of $\mathcal{N}=8$ supergravity has a similar form
 \begin{equation}
 \label{MHVsugra}
 \mathcal{M}_n^{\rm MHV}=\frac{M_{n}(1^-,2^-,3^+,\cdots,n^+)}{\langle 12\rangle^8}\,
\delta^{(16)}\Biggl(\sum_{j=1}^n\lambda^{\alpha}_j\eta_j^a\Biggr)\,,
 \end{equation}
where $M_{n}(1^-,2^-,3^+,\cdots,n^+)$ is a tree-level MHV
pure-graviton amplitude.

It is natural to expect that superamplitudes for theories with fewer
supersymmetries are simply given by subamplitudes of the maximal
theory.  Indeed as shown in ref.~\cite{SuperSum}, by grouping
R-symmetry indices, one obtains amplitudes for the gauge multiplet of
lower supersymmetric theories.  The most trivial example is that the
amplitudes of pure non-supersymmetric Yang-Mills theory are just the
pure gluon amplitudes of $\NeqFour$ super-Yang-Mills theory, since all
other states decouple from these amplitudes. In fact, by making
appropriate projections, one can even obtain QCD amplitudes with
quarks from $\NeqFour$ tree amplitudes~\cite{DixonHenn}, leading
to compact forms of QCD amplitudes.

As a simple example, the MHV tree amplitudes for the minimal gauge
multiplets of ${\cal N}<4$ super-Yang-Mills theory are given
by~\cite{SuperSum}
\begin{equation}
{\cal A}^{\rm MHV}_n(1, 2, \ldots, n)=
\frac{ \prod_{a=1}^{\cal N}\delta^{(2)}( Q^a) }
{\prod_{j=1}^n\langle j ~(j+1)\rangle}
\,\,
\Biggl(\sum_{i<j}^n \spa{i}.{j}^{4-{\cal N}} 
\prod_{a={\cal N}+1}^{4} \eta_i^a  \eta_j^a\Biggr) \,,
\label{MHVLessSuperAmplitude}
\end{equation}
with ${\cal N}$ counting the number of supersymmetries,
$Q^a=\sum_{i=1}^n \lambda_i \eta_i^a$, and $n\ge 3$. As we discuss
below, these can be used as input building blocks to construct the
integrands of loop superamplitudes. For a recent discussion on ${\cal
  N}<4$ superamplitudes, see~\cite{Elvang2011fx}.

The parity conjugate $\overline{\rm MHV}$ superamplitude is simply
given by exchanging $\lambda_i^\alpha\leftrightarrow
\tilde{\lambda}^{\dot{\alpha}}_i$ and $\eta_i^a\leftrightarrow
\tilde{\eta}_{ia}$. To obtain the component amplitude from the
$\overline{\rm MHV}$ superamplitude, one can perform a Grassmann
Fourier transformation to convert it back to the $\eta$
representation~\cite{FreedmanUnitarity}.  Starting from the
three-point MHV and $\overline{\rm MHV}$ amplitude, one can obtain the
higher-point amplitudes in superspace via the MHV vertex expansion
constructed by Cachazo, Svr\v{c}ek and Witten (CSW)~\cite{CSW} or the
on-shell recursion relations of Britto, Cachazo, Feng, and
Witten (BCFW)~\cite{BCFGeneralized,BCFW}. The supersymmetric extension
of the former approach is given in
refs.~\cite{GGK,FreedmanGenerating,FreedmanUnitarity,SuperSum} while
that of the latter is given in refs.~\cite{AHCKGravity,
RecentOnShellSuperSpace, DualConformalTree, Drummond2008cr}.

The expression for the N$^m$MHV superamplitude constructed via the CSW
construction is given by
\begin{equation}
{\cal A}^{{\rm N}^m{\rm MHV}}_n = i^m
\sum_{\rm CSW~graphs}\int \Bigl[ \prod_{j=1}^{m} {d^4\eta_j\over P^2_j}\Bigr]
 {\cal A}^{\rm MHV}_{(1)} {\cal A}^{\rm MHV}_{(2)}\cdots 
  {\cal A}^{\rm MHV}_{(m)}{\cal A}^{\rm MHV}_{(m+1)}\,,
\label{CSWequation}
\end{equation}
where the integral is over the $4m$ internal Grassmann parameters
($d^4\eta_j \equiv \prod_{a=1}^4d\eta_j^a$) associated with the
internal legs, and each $P_j$ is the momentum of the $j$'th internal
leg of the graph. To each vertex one associates an on-shell MHV
superamplitude, and the holomorphic spinor $\lambda_{P_j}$, also
denoted as $|\Pflat_j\rangle$, associated to an internal leg is
constructed from the corresponding off-shell momentum via
$\Pflat_j = P_j  - r P_j^2/2 P_j \cdot r$, where $r$ is a null reference 
momentum~\cite{K_proj}.

Alternatively, the $n$-point N$^m$MHV tree level superamplitudes
can be obtained via the super-BCFW recursion relations. An advantage
of this approach is that the final form of the amplitude can be given
so that dual conformal symmetry is
manifest~\cite{DualConformalTree}.\footnote{The amplitudes are
  actually dual superconformal covariant, not invariant. This
  covariance can be translated into an invariance, by considering the
  Yangian generators which are obtained by combining ordinary and
  dual superconformal generators~\cite{Drummond2009fd,BeisertReview}. }
This
symmetry~\cite{DualConformal, RecentOnShellSuperSpace} is a conformal
symmetry defined in the dual space $x$, which is related to the
external momenta via
\begin{equation}
x_i-x_{i+1}=p_i\,,
\label{xdefine}
\end{equation}
where $x_{n+1} \equiv x_1$.  (Further details on dual (super)conformal
symmetry may be found in the chapters of the present review by
Drummond and by Henn~\cite{DrummondReview,HennReview}.)  Using 
super-BCFW recursion relations, Drummond and Henn were able to derive
all N$^m$MHV tree amplitudes in terms of dual conformal
invariants in ref.~\cite{Drummond2008cr}. For example, the $n$-point
NMHV amplitude is given as
\begin{equation}
\mathcal{A}^{\rm NMHV}_n(1,2,\cdots,n)=
  \mathcal{A}^{\rm MHV}_n\sum_{1<s,t<n}R_{n;st}\,,
\end{equation}
where the function $R_{n;st}$ is a dual conformal invariant
function. The specific form, and its N$^m$MHV generalization
may be found in ref.~\cite{Drummond2008cr,HennReview}.

The tree amplitudes obtained from the BCFW or CSW procedure in general
contain spurious poles which cancel between different contributions.
The spurious poles complicate the merging procedure of different
cuts. However, if one works in the reverse direction starting from an
ansatz, one can use numerical analysis to solve for coefficients of
the ansatz for the amplitude.  In this way, complications from the
spurious poles can be easily avoided.  (At one loop see also 
the discussion in Britto's chapter of the review~\cite{BrittoReview}.)

\subsection{Loop superamplitudes}

We now turn to the construction of four-dimensional unitarity cuts in
superspace.  The sum over states crossing the unitarity cuts can be
expressed simply as an integration over the $\eta^a$ parameters of the
cut legs, since this gives the states properly on either side of the
cut.  The generalized $\NeqFour$ supercut is then given by
\begin{equation}
{\cal C} = 
\int \Bigl[ \prod_{i=1}^{k} {d^4\eta_i}\Bigr]
 {\cal A}^\tree_{(1)}
  {\cal A}^\tree_{(2)}
  {\cal A}^\tree_{(3)} \cdots 
   {\cal  A}^\tree_{(m)}\,,
\label{SuperCut}
\end{equation}
where ${\cal A}_{(j)}^{\rm tree}$ are the tree superamplitudes
connected by $k$ on-shell cut legs. These cuts then constrain the
amplitude, which are now functions in the
on-shell superspace. For four and higher points the tree amplitude
${\cal A}_{(j)}^{\rm tree}$ is always proportional to a supermomentum
delta function\footnote{The exception is the three-point \MHVbar{}
  superamplitude where the supermomentum conservation is not
  manifest~\cite{KorchemskyOneLoop}.}.  Using the identity
$\delta(A)\delta(B)=\delta(A+B)\delta(B)$, this implies that all such
cuts are proportional to an overall supermomentum delta
function~\cite{KorchemskyOneLoop}. As discussed in
\sect{BasicUnitaritySection}, a spanning set of cuts can be constructed
using four and higher-point tree amplitudes.
This then implies that any loop amplitude with four or more legs must
be proportional to an overall supermomentum conservation delta
function. One interesting consequence is that this overall delta
function implies that at least four would be powers of loop momenta in
the numerators of each diagram are external momenta.  Simple power
counting then implies the well known~\cite{Mandelstam} ultraviolet
finiteness of all $\NeqFour$ super-Yang-Mills amplitudes~\cite{SuperSum}.

There are a number of ways of evaluating supersums in unitarity
cuts~\cite{KorchemskyOneLoop, AHCKGravity, FreedmanUnitarity,
SuperSum}.  Here we describe two such approaches for organizing the
integration over the Grassmann parameters.  In the first way, the
fermionic delta functions can be used to localize the integration, so
that the evaluation of the supersum amounts to solving a subset of
linear equations~\cite{KorchemskyOneLoop,SuperSum}. In a second
complementary approach one uses ``index diagrams'' to track the
contributions~\cite{SuperSum}. This latter approach leads to a simple
algorithm for reading off the contribution of the entire
supermultiplet from the purely gluonic ones or for reducing supersymmetry and
was used in the construction of the complete four-loop four-point
amplitude of $\NeqFour$ super-Yang-Mills theory~\cite{Neq44np}.

\subsection{The supersum as a system of linear equations}
\label{LinearSystemSection}

We now consider in more detail the evaluation of supersums. 
First consider the approach based on solving a system of
equations. This task is straightforward
if the product of tree amplitudes
is written in a form that depends on the $\eta^a$s only through
fermionic delta functions. The MHV vertex expansion offers
a simple way to accomplish this~\cite{CSW}.
The arguments of the delta functions then
automatically form a set of linear equations that constrain the
$\eta^a$s. 

Simple counting shows that after the overall supermomentum
conservation constraint is extracted, the number of equations
appearing in cuts of MHV loop amplitudes equals the number of
integration variables.  For such cuts the integrals are completely
localized by delta functions, and one simply computes the Jacobian of
the linear equations.

For N$^m$MHV amplitudes, the number of constraints will be
larger than the number of integration variables. For this case, one
proceeds by solving some judiciously chosen subset of the
supermomentum constraints and substituting the result into the remaining
fermionic delta functions. However, some choices of solving subsets of the
constraints may lead to results that obscure symmetries of the
amplitude. One simple helpful strategy is to choose the constraints with
as few external momenta as possible.

\subsubsection{Example: The one-loop four-point amplitude.}

To illustrate these ideas we turn to a few simple examples.  As a
first example, we revisit the computation of the one-loop four-point
amplitude of $\NeqFour$ super-Yang-Mills theory. Consider the
$s$-channel supercut of the one-loop four-point superamplitude shown
in \fig{TwoParticleCutFigure}(a). The supersum is obtained from the
Grassmann integrals,
\be
{\cal C}_s=
 \int d^4\eta_{l_1}\int d^4\eta_{l_3}  \,
{\cal A}_4^{\rm MHV}(-l_1,1,2, l_3)
{\cal A}_4^{\rm MHV}(-l_3,3,4, l_1)\,,
\label{OneLoopCutSuperExample}
\ee
where the MHV superamplitudes are given by relabeling
\eqn{MHVSuperAmplitude} for $n=4$.  The $\eta$ integration acts on the
two supermomentum delta functions contained in the tree
superamplitudes, 
\be
\hskip -.5 cm 
\delta^{(8)}\Bigl(\lambda_{l_3}^\alpha\eta_{l_3}^a
             -\lambda_{l_1}^\alpha\eta_{l_1}^a
             +\lambda_{1}^\alpha\eta_{1}^a+\lambda_{2}^\alpha\eta_{2}^a \Bigr) \;
\delta^{(8)}\Bigl( \lambda_{l_1}^\alpha\eta_{l_1}^a
             -\lambda_{l_3}^\alpha\eta_{l_3}^a
             +\lambda_{3}^\alpha\eta_{3}^a+\lambda_{4}^\alpha\eta_{4}^a \Bigr) \,.
\ee
Adding the argument of the first delta function to the second one, one
obtains an overall supermomentum conservation delta function that
can be extracted outside of the Grassmann integral, 
\be
\hskip -.5 cm 
{\cal C}_s=-\delta^{(8)}\Bigl(\sum_{i=1}^{4}\lambda_{i}^\alpha\eta_{i}^a\Bigr)
\frac{ \int d^4\eta_{l_1}d^4\eta_{l_3}\delta^{(8)}\Bigl(\lambda_{l_3}^\alpha\eta_{l_3}^a
             -\lambda_{l_1}^\alpha\eta_{l_1}^a
             +\lambda_{1}^\alpha\eta_{1}^a+\lambda_{2}^\alpha\eta_{2}^a \Bigr)}
{ \spa{l_1}.1 \spa1.2\spa2.{l_3}\spa{l_3}.{l_1}
\, \spa{l_3}.3 \spa3.4 \spa4.{l_1} \spa{l_1}.{l_3}}\,. 
\label{FourPtGrassmannInt}
\ee
Choosing an explicit value for the R-symmetry index, for example
$a=1$, the fermionic integration becomes,
\be
\int d\eta_{l_1}^a d\eta_{l_3}^a\delta^{(2)}\Bigl(\lambda_{l_3}^\alpha\eta_{l_3}^a
             -\lambda_{l_1}^\alpha\eta_{l_1}^a
             +\lambda_{1}^\alpha\eta_{1}^a+\lambda_{2}^\alpha\eta_{2}^a \Bigr)=-\langle l_3l_1\rangle\,,
\ee 
which may be read off using the form of the delta function displayed
in \eqn{DeltaSpinor}.
Similarly, the cases $a=2,3,4$ give the identical factor.
Thus, the entire fermionic integration simply gives a factor of
$\langle l_3l_1\rangle^4$.

Alternatively, the fermionic delta function inside the 
Grassmann integral in \eqn{FourPtGrassmannInt} 
can be viewed as enforcing the constraints,
 \be
\lambda_{l_1}^\alpha\eta_{l_1}^a-\lambda_{l_3}^\alpha\eta_{l_3}^a
=\lambda_{1}^\alpha\eta_{1}^a+\lambda_{2}^\alpha\eta_{2}^a \,. 
\ee 
There are a total of eight conditions which matches the eight integration
variables $\eta_{l_1}^a,\eta_{l_3}^a$. Thus, the delta functions
completely fix the integration and one obtains the Jacobian
of the matrix of the coefficients of the linear equations,
\begin{eqnarray}
J &=&  \det{}^{\! 4} \left|
\begin{array}{cc}
  \lambda_{l_1}^1 & -\lambda_{l_3}^1 \\
  \lambda_{l_1}^2 & -\lambda_{l_3}^2 
\end{array}
\right| = 
\spa{l_1}.{l_3}^4 \,.
\end{eqnarray}

Either way, the $s$-channel cut of the superamplitude is
\begin{eqnarray}
{\cal C}_s&=&
-\frac{\delta^{(8)}\Bigl(\sum_{i=1}^n\lambda_i^\alpha\eta_i^a\Bigr)
\, \spa{l_1}.{l_3}^4}
{ \spa{l_1}.1 \spa1.2\spa2.{l_3}\spa{l_3}.{l_1}
\, \spa{l_3}.3 \spa3.4 \spa4.{l_1} \spa{l_1}.{l_3}}\,. 
\end{eqnarray}
Other than the replacement of a factor of $\spa1.2^4$ by the overall
supermomentum delta function, this is the same expression as arrived
at by using components as in \eqn{SCutSusyA}.  
Thus, following the same steps as for components one obtains
\begin{eqnarray}
{\cal C}_s={\cal A}_4^{\rm tree}(1,2,3,4)\, \frac{-ist}{(p-k_1)^2(p+k_4)^2}\,.
\label{D4CutResult}
\end{eqnarray}
Putting back the cut propagators and comparing with the $t$-channel
cut, one arrives at the four-point one-loop superamplitude,
\begin{equation}
{\cal A}_4^{\rm1-loop}= -st{\cal A}^{\rm tree}_4 I_4(s,t) \,,
\end{equation}
where the box integral $I_4(s,t)$ is defined in \eqn{BoxIntgegral}.
This result is in agreement with the component analysis of
\sect{ComponentsSampleSection}.

\subsubsection{Example: The two-loop four-point amplitude using 
the MHV expansion.}
\label{ruleMHVSubSection}

\begin{figure}[t]
\centerline{\epsfxsize 1.9 truein \epsfbox{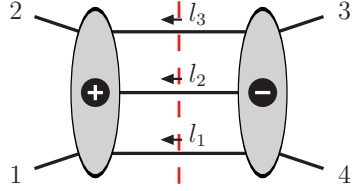}}
\caption{A three-particle supercut for the four-point 
amplitude.  This cut contribution contains one MHV and one \MHVbar{}
superamplitude.  The ``$+$'' indicates the MHV amplitude, while 
the ``$-$'' indicates the \MHVbar{} one.
 \label{ThreeCutSuperSumMHVbarFigure}}
\end{figure}

\begin{figure}[t]
\centerline{\epsfxsize 1.9 truein \epsfbox{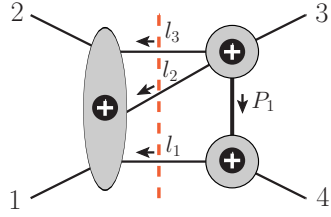}}
\caption{The same cut as in \fig{ThreeCutSuperSumMHVbarFigure},
except that \MHVbar{} tree amplitude on the right is expanded in 
terms of MHV amplitudes.  We display only one of the terms here.
The thick line labeled by $P$ marks the internal propagator
used in the MHV expansion. 
 \label{ThreeCutSuperSumFigure}}
\end{figure}

As a more sophisticated example, consider the three-particle cut of
the two-loop four-gluon amplitude displayed in
\fig{ThreeCutSuperSumMHVbarFigure}. This cut contains both MHV and
\MHVbar{} amplitudes.  This cut was evaluated in components in
refs.~\cite{BRY,BDDPR} and in superspace in
ref.~\cite{FreedmanUnitarity,SuperSum}. The supercut of
\fig{ThreeCutSuperSumMHVbarFigure} may be conveniently evaluated using
a Grassmann Fourier transform.  Instead, for illustrative purposes
here we expand the \MHVbar{} tree amplitude using the MHV vertex
expansion~\cite{CSW}. (See the chapter in this review~\cite{BSTReview} 
giving further details on the MHV expansion.)  This
procedure provides a general means for evaluating any cut not
involving \MHVbar{} three-point tree amplitudes.\footnote{If
  three-point \MHVbar{} tree amplitudes are present in a given cut, we
  can either absorb them into cuts with higher-point trees, or apply
  Fourier transform techniques~\cite{FreedmanUnitarity,SuperSum}.}
The MHV vertex expansion reduces any generalized cut to cuts involving only MHV
amplitudes.  We thus expand the \MHVbar{} tree amplitude on the right
side of the cut in \fig{ThreeCutSuperSumMHVbarFigure}.  One of the
obtained terms in the MHV expansion is shown
\fig{ThreeCutSuperSumFigure}.  Here we will describe how to obtain the
result for this specific supercut contribution; the remaining ones are
similar.

For the cut in \fig{ThreeCutSuperSumFigure}, for a fixed R-symmetry
index $a$ we have three supermomentum constraints from each of the
three MHV amplitudes,
\begin{eqnarray}
\nonumber&&\delta^{(2)}(\lambda_1^\alpha\eta_1^a+\lambda_2^\alpha\eta_2^a
-\lambda_{l_1}^\alpha\eta_{l_1}^a-\lambda_{l_2}^\alpha\eta_{l_2}^a
-\lambda_{l_3}^\alpha\eta_{l_3}^a)\\
\nonumber&&\times\delta^{(2)}(\lambda_3^\alpha\eta_3^a + \lambda_{P_1}^\alpha\eta_{P_1}^a
+\lambda_{l_2}^\alpha\eta_{l_2}^a +\lambda_{l_3}^\alpha\eta_{l_3}^a)\\
&& \times \delta^{(2)}(\lambda_4^\alpha\eta_4^a
+\lambda_{l_1}^\alpha\eta_{l_1}^a - \lambda_{P_1}^\alpha \eta_{P_1}^a)\,.
\label{initial_const}
\end{eqnarray}
In this expression $l_1, l_2$ and $l_3$ refer to the momenta of 
the three cut lines. The spinors labeled by $P_1$ satisfy 
$\lambda_{P_1} \tilde \lambda_{P_1} = \Pflat_1 = P_1 - r P_1^2/2P_1\cdot r$,
with $r$ a null reference momentum and $P_1 = l_1 + k_4$, as 
arises in the MHV expansion~\cite{CSW,K_proj}.
We first extract the overall supermomentum conservation delta function
and leave behind the second and third delta function. These
enforce the constraints,
\begin{eqnarray}
-\lambda_{P_1}^\alpha \eta_{P_1}^a
\hphantom{+\lambda_{l_1}^\alpha\eta_{l_1}^a}
-\lambda_{l_2}^\alpha \eta_{l_2}^a -\lambda_{l_3}^\alpha\eta_{l_3}^a
 &=&\lambda_3^\alpha\eta_3^a \,,
\cr
+\lambda_{P_1}^\alpha\eta_{P_1}^a
-\lambda_{l_1}^\alpha\eta_{l_1}^a
\hphantom{-\lambda_{l_2}^\alpha \eta_{l_2}^a -\lambda_{l_3}^\alpha\eta_{l_3}^a}
&=&\lambda_4^\alpha\eta_4^a \,.
\end{eqnarray}
Since there are 16 equations for $4\times4=16$ variables, the
integrals are completely localized and one simply needs the Jacobian
of the matrix of coefficients,
\be 
\left( \!
\begin{array}{cccc}
-\lambda_{P_1} ^\alpha& 0 & -\lambda_{l_2}^\alpha& -\lambda_{l_3}^\alpha \cr
+\lambda_{P_1} ^\alpha& -\lambda_{l_1}^\alpha& 0 & 0
\end{array} \!\right) \,,
\ee
where each spinor $\lambda^\alpha_j$ should be thought of as a
submatrix with two rows and one column.  The Jacobian of this matrix
is just $\spa{l_1}.{\Pflat_1} \spa{l_2}.{l_3}$.
For each of the four values that the R-symmetry index takes
on we get the same result.
Thus the
Grassmann integration over the delta functions in \eqn{initial_const}
evaluates to,
\be 
(\langle l_1 \Pflat_1\rangle \langle l_2 l_3\rangle)^4 
\delta^{(8)}\Bigl(\lambda_1^\alpha \eta_1^a+\lambda_2^\alpha\eta_2^a
+\lambda_3^\alpha \eta_3^a+\lambda_4^\alpha \eta_4^a \Bigr)\,.  
\ee
To obtain the complete result, one needs to repeat the same steps for
other diagrams corresponding to the other terms in the MHV
expansion of the five-point \MHVbar{} amplitude.  Although the result
is overly complicated because of the expansion of a simple \MHVbar{}
tree amplitude in MHV diagrams, it does illustrate a general
technique for evaluating supersums in unitarity cuts.

\subsubsection{Example: The two-loop four-point amplitude using the 
N$^m\!$MHV dual conformal representation.}

As another means for evaluating generalized cuts in $\NeqFour$
super-Yang-Mills theory, one can start from the closed form N$^m$MHV
superamplitudes derived in ref.~\cite{Drummond2008cr}.  (These tree
solutions also may be in found in the chapter of this review from
Drummond~\cite{DrummondReview}.)  One proceeds by using the fermionic
delta functions to localize part of the integral, and implementing the
solution of the constraints into the remaining superfunction. The
final integrals then pick out specific pieces of the superfunction.

More concretely, consider again the tree amplitude on the right-hand-side of the
three-particle cut in \fig{ThreeCutSuperSumMHVbarFigure} as an ${\rm NMHV}$
superamplitude. The supermomentum delta functions can be used to
localize a total of two loop $\eta^a$s for each ${\rm SU(4)}$
R-index. Take them to be $l_1$ and $l_2$. The supersum can then be
rewritten as
 \begin{eqnarray}
{}&&\int \left[\prod_{i=1}^3d\eta^a_{l_i}\right]\;
\delta^{(2)}(Q^a_L)\delta^{(2)}(Q^a_R)\left(R_{3;l_1l_2}+R_{3;l_3l_2}+R_{3;l_3l_1}\right)\,,  \hskip 1 cm 
\label{SuperSumDualConf}
\end{eqnarray}
where 
\begin{eqnarray}
Q_L^{\alpha a} &=& \lambda_1^\alpha\eta_1^a + \lambda_2^\alpha \eta_2^a - 
                 \lambda_{l_1}^\alpha \eta_{l_1}^a - 
                 \lambda_{l_2}^\alpha \eta_{l_2}^a - 
                 \lambda_{l_3}^\alpha \eta_{l_3}^a\,,  \nn\\
Q_R^{\alpha a} &=& \lambda_3^\alpha \eta_3^a + \lambda_4^\alpha \eta_4^a +
                 \lambda_{l_1}^\alpha \eta_{l_1}^a +
                 \lambda_{l_2}^\alpha \eta_{l_2}^a +
                 \lambda_{l_3}^\alpha \eta_{l_3}^a\,, 
\end{eqnarray}
and the $R$s are a set of dual conformal invariant functions
mentioned in \sect{SuperTreeSubsection}.  Pulling out the overall
supermomentum delta function and leaving behind $\delta^{(2)}(Q^a_R)$,
we can rewrite \eqn{SuperSumDualConf} as
\begin{eqnarray}
{}\langle l_1l_2\rangle\delta^{(2)}\Bigl(\sum_{i=1}^4q^a_i \Bigr)
 &&\int \left[\prod_{i=1}^3d\eta^a_{l_i}\right]
\delta \Biggl(\eta^a_{l_1}+
\frac{\langle l_2 |(q^a_{l_3}+q^a_{3}+q^a_{4})}{\langle l_2l_1\rangle}\Biggr)
\label{Localize}
 \\
&& \hskip - .3 cm 
\times\delta\Biggl(\eta^a_{l_2}+\frac{\langle l_1 |(q^a_{l_3}+q^a_{3}+q^a_{4})}{\langle l_1l_2\rangle}\Biggr)
(R_{3;l_1l_2}+R_{3;l_3l_2}+R_{3;l_3l_1})\,.
 \nn
\end{eqnarray}
Thus, the delta functions simply localizes the
$\eta^a_{l_1},\eta^a_{l_2}$ integral and we obtain
\begin{eqnarray}
{}\delta^{(2)}\Bigl(\sum_{i=1}^4 q^a_i \Bigr) \int d\eta^a_{l_3}
\langle l_1l_2\rangle\left(R_{3;l_1l_2}+R_{3;l_3l_2}+R_{3;l_3l_1}\right)
 \Bigr|_{\rm sol}, 
\end{eqnarray}
where $(\cdots +  R_{3;l_3l_1})|_{\rm sol}$ signifies
that  all
$\eta^a_{l_1},\eta^a_{l_2}$ in the functions $R_{n;st}$ have been set
to the solution of the fermionic constraints, 
\begin{equation}
\eta^a_{l_1}=-\frac{\langle l_2 |(q^a_{l_3}+q^a_{3}+q^a_{4})}{\langle
  l_2l_1\rangle}\,,
\hskip 1 cm \eta^a_{l_2}=-\frac{\langle l_1
  |(q^a_{l_3}+q^a_{3}+q^a_{4})}{\langle l_1l_2\rangle} \,.  
\end{equation}
The
final $\eta^a_{l_3}$ integration then simply picks out terms
proportional to $q^a_{l_3}$ in the final expression.  

In all of the above examples,
after pulling out the overall supermomentum delta function, all
fermionic variables are integrated away in the final expression. This
property is unique to MHV loop amplitudes. For N$^m$MHV
amplitudes, there will be $4m$ fermionic variables left after
performing all Grassmann integrations. A convenient way of organizing
these variables is given in ref.~\cite{KRV}. For planar amplitudes in
the $\NeqFour$ theory, the presence of dual conformal symmetry implies
that these variables can be organized into the invariant function
$R_{n;st}$ and its generalizations.

\subsection{Supersums as SU(4) index diagrams }

As manifest in the above examples, evaluations of the supersums
appearing in cuts of $\NeqFour$ super-Yang-Mills theory boil down to
evaluating the $\eta_a$ integration for a fixed SU(4) R-index, and
taking the result to the fourth power. In this section, we explain
this structure by keeping track of the different contributing states
using a diagrammatic language to track the R-symmetry
indices~\cite{SuperSum}. This leads to a surprisingly simple algorithm
for essentially reading off contributions to the supercuts starting
from non-supersymmetric purely gluonic cuts.  This algorithm provided
a simple means to evaluate the supersums in 
the large number of generalized cuts providing 
cross checks in the calculation of the four-loop four-point amplitudes
of $\NeqFour$ super-Yang-Mills theory~\cite{Neq44np} and $\NeqEight$
supergravity~\cite{GravityFour}.  Because individual states are
tracked, the diagrammatic language also easily allows us to reduce the
supersymmetry.

\begin{figure}[th]
\centerline{\epsfxsize 1.5 truein \epsfbox{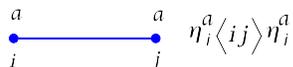}}
\caption[a]{\small  For an MHV amplitude the shaded (blue) ``index line''  connecting leg $i$ to leg $j$ represents $\langle q_i^a \,
q_j^a\rangle$.  }
\label{MHVruleFigure}
\end{figure}
%
\subsubsection{Structure of index diagrams.}

\begin{figure}[th]
\centerline{\epsfxsize 5.6 truein \epsfbox{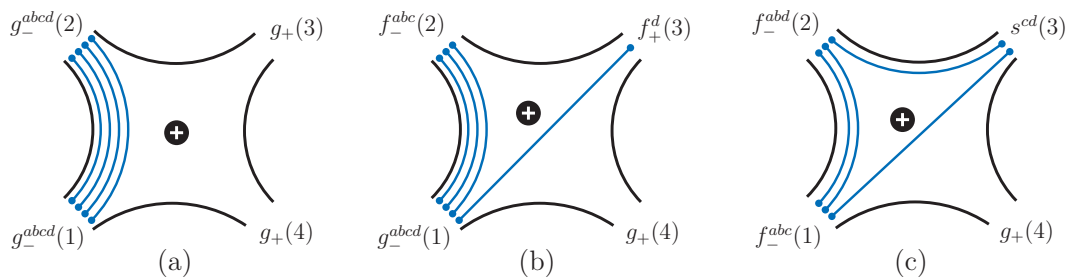}}
\caption[a]{\small Examples of R-symmetry index diagrams for specifying
different component amplitudes. Case (a) is a pure gluon amplitude,
case (b) a two-fermion two-gluon amplitude, and case (c) a
two-fermion one-scalar one-gluon amplitude. The black outer lines connect
the external legs and the shaded (blue) lines are the R-symmetry index lines.
A white ``$+$'' and ``$-$'' on a black background indicates whether
the amplitude is MHV or \MHVbar{}.
}
\label{SampleMHVTreesFigure}
\end{figure}

To see how the index diagrams arise, we first look at tree amplitudes.
Up to the fact that all fields in the $\NeqFour$ multiplet are in
the adjoint color representation, one sees that the SU(4) R symmetry is
analogous to the SU($N_{\!f}$) flavor symmetry of QCD. Much like QCD, where
flavor lines must be conserved within the amplitude, one can
construct R-symmetry conserving lines for $\NeqFour$ super-Yang-Mills
theory. This is most
transparent for  MHV tree amplitudes, where the supermomentum delta
function can be considered as four copies of
\begin{equation}
\langle q^a_iq_j^a\rangle\equiv\eta^a_i\langle ij\rangle\eta_j^a\,.
\end{equation}  
The pair of external legs $(i,j)$ associated with each R index can be
different.  Pictorially, we can represent the supermomentum product by
a shaded (blue) line connecting legs $i$ and $j$, as displayed
\fig{MHVruleFigure}.  We will call this object an ``index line''.
Since different component amplitudes corresponds to different R-index
structures for the external legs, in terms of index diagrams, this
translate into different ``paths" for the index lines. We illustrate
the different component amplitudes of the four-point MHV amplitudes in
terms of index diagrams in \fig{SampleMHVTreesFigure}. Note that the
pure gluon amplitudes corresponds to four index lines following the
same path as in \fig{SampleMHVTreesFigure}(a).  Amplitudes with
fermions or scalars correspond to cases where index lines follow
different paths as illustrated in \fig{SampleMHVTreesFigure}.  
We can also define index lines for \MHVbar{} amplitudes by
switching the role of positive and negative helicity helicity spinors.

The index line language generalizes straightforwardly to any tree
amplitude via the MHV expansion~\cite{CSW}.  In this expansion we
obtain sums of products of on-shell tree amplitudes, but with shifted
momenta.  To obtain an index-diagram representation we simply dress
each MHV component amplitude by the index diagrams.  This diagrammatic
representation also extends straightforwardly to loops.  To do so we
expand any component tree amplitude appearing in the cuts that is not
MHV or \MHVbar{} into such amplitudes using the MHV expansion.  If
both MHV and \MHVbar{} tree amplitudes are present in the cuts, the
index lines can either cross, terminate or begin at the cuts.  The details are
given in ref.~\cite{SuperSum}.

\begin{figure}[t]
\centerline{\epsfxsize 6 truein \epsfbox{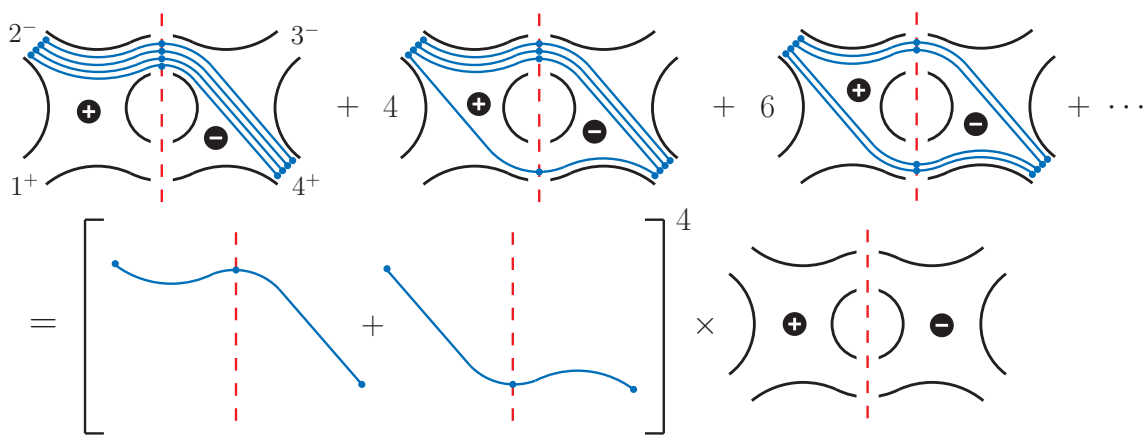}}
\caption[a]{\small A unitarity cut of the four-gluon amplitude
$A^\oneloop_4 (1^+,2^-,3^-,4^+)$, involving one MHV and one \MHVbar{}
superamplitude.  The top-left diagram represents a gluon loop, the
top-central diagram represents the four contributions in a fermion
loop, and the top-right diagram represents the six scalar state
contributions.  The ellipsis denote that four more fermion-loop and
one more gluon-loop contributions are suppressed.  
}
\label{OneLoopMHVbarExampleFigure}
\end{figure}

To illustrate the basic idea, we use the simple case of a cut of the
one-loop four-point $\NeqFour$ amplitude, shown in
\fig{OneLoopMHVbarExampleFigure}. The left side of the cut is chosen
to be MHV and right side is \MHVbar{}. In this case, the SU(4)
R-symmetry index lines are continuous through the cut.  The different
diagrams in the top line of \fig{OneLoopMHVbarExampleFigure} correspond to
the different states crossing the cuts, the two hidden in the ellipsis
are horizontal flips of the first two shown.  The combinatoric factors
in front of each diagram are the distinct ways of obtaining the same
diagram, tracking the four distinct SU(4) labels.

As shown in \fig{OneLoopMHVbarExampleFigure}, the sum over the
diagrams can be interpreted as a product over the four R-symmetry indices,
depicted as a fourth power.  The pure gluon configuration then
corresponds to all four index lines taking the same ``path''. Mixed
paths then correspond to configurations with scalars and fermions.
This structure, in fact, holds for any cut composed of MHV and
\MHVbar{} amplitudes.  Since the MHV expansion allows us to express
{\it any} cut composed of tree amplitudes in terms of MHV and
\MHVbar{} amplitude, for $\NeqFour$ super-Yang-Mills amplitudes all cuts
are then given by a sum of terms, each with a numerator raised
to the fourth power.
 
\subsubsection{$\NeqFour$ loop amplitudes from pure gluon amplitudes.}

The one-loop discussion extends straightforwardly to
multiloops~\cite{SuperSum}. 
By analyzing these index diagrams a simple pattern emerges: in 
the $\NeqFour$
theory for each configuration of MHV and \MHVbar{} amplitudes in a
given cut, the supersums' contributions for external gluons are always
proportional to the fourth power of a sum of terms,
\begin{eqnarray}
	(A+B+C+\cdots)^{ 4}\,,
	\label{NeqFourSupersum}
\end{eqnarray}
where the summands $A, B, C,\dots$ represent different spinorial
numerator factors encoded by the R-symmetry index diagrams. On the
other hand, the pure gluon contribution takes the form
\begin{eqnarray} 
	(A^4+B^4+C^4+\cdots) \,.
	\label{NeqZeroSupersum}
\end{eqnarray}
Comparing the two, we see that the supersum result
(\ref{NeqFourSupersum}) is identical to the pure gluon result 
(\ref{NeqZeroSupersum}), except the fourth power appears on the sum of
terms instead of individual terms.  Thus, we can read off the
contributions of the entire $\NeqFour$ multiplet directly from the
pure gluon contributions, up to relative signs.  The relative signs
between the terms in \eqn{NeqFourSupersum} can be tracked by including
Grassmann parameters.  The details of this procedure are presented in
ref.~\cite{SuperSum}.

\subsubsection{Amplitudes with fewer supersymmetries from $\NeqFour$
ones.}

As a further application, the pure gluon result can also be used to
generate cut results for theories with fewer supersymmetries. This is
straightforward for theories that can be viewed as a truncation of
$\NeqFour$ Yang-Mills, {\it i.e.,} their action can be obtained from
the $\NeqFour$ theory by retaining a subset of fields, forming a
SU($\mathcal{N}$) representation with $\mathcal{N}<4$, and setting the
remaining ones to zero. This corresponds to $\mathcal{N}=1,2$
super-Yang-Mills theories.

To see how to effectively implement the truncation to theories with 
fewer supersymmetries, we break the SU(4) R-symmetry group into
SU($\mathcal{N}$)$\times$SU($4-\mathcal{N}$)$\times$U(1). The on-shell
states of the lower supersymmetric theories transform under the
SU($\mathcal{N}$) R-symmetry group, but behave as singlets under the
remaining SU($4-\mathcal{N}$). In terms of index diagrams, the singlet
criteria translates to grouping the index lines that carry the
SU($4-\mathcal{N}$) together, {\it i.e.,} these index lines must take
the same path. This guarantees that the states crossing the cuts are
either completely independent of the SU($4-\mathcal{N}$) indices, or
its dependence is proportional to the SU($4-\mathcal{N}$) invariant
tensor $\epsilon_{(4-\mathcal{N})\cdots 4}$.

For an example, consider $\mathcal{N}=2$ super-Yang-Mills theory. We
take the contributing index diagrams to be those where index lines
labeled by $3$ and $4$ follow the same path. For external gluons, this
gives the following numerator in any MHV cut contributions,
\begin{equation}
(A+B+C+\dots)^2(A^2+B^2+C^2+\dots)\,,
\label{NeqTwoSupersum}
\end{equation}
where $A, B, C$ represent the same terms as in \eqn{NeqFourSupersum},
and the squares $A^2, B^2, C^2$ are a consequence of the requirement
that two indices are always grouped together in the diagrams.    

A straightforward generalization to any number of supersymmetries 
then gives~\cite{SuperSum}
\begin{eqnarray} 
(A+B+C+\dots)^{\cal N}\times (A^{4-{\cal N}}+B^{4-{\cal N}}+
C^{4-{\cal N}}+\dots)\,,\quad {\cal N}<4\,,
	\label{GenSupersum}
\end{eqnarray}
which holds for ${\cal N}=0,1,2,3$.  (The ${\cal N}=3$ case is
identical to the $\NeqFour$ super-Yang-Mills case
\eqn{NeqFourSupersum}, as expected from the well known equivalence of
these theories~\cite{N3N4Equiv}.) In \eqn{GenSupersum}, the
first factor corresponds the supersymmetric summation over index lines
with ${\cal N}$ independent R-symmetry indices, and the second factor
corresponds to the truncation of index diagrams, where 
$4-{\cal N}$ indices are always grouped together.

\section{Six-dimensional $\mathcal{N}=(1,1)$ on-shell superspace}

In many cases, it is necessary to work in higher dimensions, either
for regularization purposes or to study properties of theories in
$D>4$ dimensions.  As mentioned above, because the $\NeqFour$
super-Yang-Mills amplitudes are infrared divergent, either dimensional
or a massive regularization is needed.  At one-loop a theorem
guarantees that four-dimensional cuts naively extended to dimensional
regularization will capture all contributions.  However, no such
theorem yet exists at higher loops.  To ensure that no terms are
dropped even in four dimensions, one should compute the cuts in $D > 4$
dimensions or with masses. A simple way to generate massive amplitudes
is to again work in higher dimensions, but treat the extra-dimensional
momenta as masses which are not integrated over.  Of course, one would
also like to use a helicity formalism in higher dimensions.
Fortunately, as noted earlier, a spinor-helicity formalism does exist in
six dimensions~\cite{CheungOConnell}, along with an on-shell maximal
superspace~\cite{DHS}. (For a general discussion of on-shell
superspaces see ref.~\cite{Boels}.)  Here we will describe the
six-dimensional helicity formalism, focusing on its application to the
construction of loop amplitudes in a manner manifestly consistent
with unitarity~\cite{SixD,SixDTwo}.

\subsection{Tree-level superamplitudes}

In six dimensions where the Lorentz group is
SO(5,1)$\sim$SU$^*$(4),\footnote{Here the ``$*$'' indicates it is in the
  pseudoreal representation.} a vector is in the anti-symmetric
representation of SU$^*$(4), $p^{AB}=-p^{AB}$, where $A,B=1,\cdots,4$. A
null vector can then be written as a spinor product in the bispinor
notation as
\begin{equation}
p^2=0\rightarrow p^{AB}=\lambda^{Aa}\lambda^{B}_a\,,
\label{SixDMom}
\end{equation}
where $\lambda^{Aa}$ is a six-dimensional chiral spinor, carrying a
local SU(2) index, $a$. The momentum $p^{AB}$ is invariant under the
local SU(2) transformation, which gives the correct counting for the
degrees of freedom for a massless six-dimensional vector,
$4\times2-3=5$. However, since the momentum is also invariant under
the global part of the SU(2) transformation, this symmetry is
isomorphic to the chiral part of the little group. From now on we will
identify the two.  A gluon polarization vector can then be readily
expressed as~\cite{CheungOConnell}
\begin{eqnarray}
  \epsilon^\mu_{a \dot a}(p,k) = \frac{1}{\sqrt{2}} 
              \langle p_a | \sigma^\mu|k_b\rangle 
             (\langle k_b | p^{\dot a}])^{-1}
                      =  
             (\langle p^a | k_{\dot b}])^{-1}
     \frac{1}{\sqrt{2}} [ k_{\dot b} | \tilde\sigma^\mu|p_{\dot a} ] 
	      \,,
\label{polar}
\end{eqnarray}
where $k$ is an arbitrary null vector and we used the bra-ket notation
for the chiral $\lambda^{Aa}$and anti-chiral spinor
$\tilde{\lambda}_{A\dot{a}}$ respectively. Note that the polarization
vector carries the full six-dimensional little-group indices
SO(4)$\,\sim\,$SU(2)$\times$SU(2), {\it i.e.} the on-shell states of
the gauge field are non-chiral. This is an important distinction from
the four-dimensional descendant, where the vector polarization states
are chiral.

For the maximal supersymmetric theory in six dimensions, an on-shell
superspace has been constructed~\cite{DHS} that has many similarities with the
four-dimensional version~\cite{Nair}. The six-dimensional on-shell
states include four vectors, four scalars, and eight
fermions. These states can be packaged into a single scalar
superfield, with the supercoordinates $\eta_a,\tilde{\eta}_{\dot{a}}$
each carrying a little-group index. The expansion of the superfield in
terms of component fields takes
the form,
\begin{eqnarray}
\Phi(\eta,\tilde{\eta}) &=&
  \phi 
  + \chi^a \eta_a 
  + \phi'(\eta)^2 
  + \psi_{\dot a}\tilde{\eta}^{\dot a} 
  + g^a\,_{\dot a}\eta_a\tilde{\eta}^{\dot a}
  + \tilde{\psi}_{\dot a}(\eta)^2\tilde{\eta}^{\dot a} \nn\\
  && \null
  + \phi''(\tilde{\eta})^2
  + \tilde{\chi}^a\eta_a(\tilde{\eta})^2
  + \phi'''(\eta)^2(\tilde{\eta})^2 \,.
\end{eqnarray}
The projection down to four-dimensional states is done by noting that the
four-dimensional U(1) little group emerges as the diagonal U(1) of the
six-dimensional SU(2)$\times$SU(2). 

Using six-dimensional spinors, we define the supermomentum as
\begin{equation}
q^A=\lambda^{Aa}\eta_a,\; \tilde{q}_B=
    \tilde{\lambda}_{B\dot{b}}\tilde{\eta}^{\dot{b}} \,.
\end{equation}  
These variables give fairly
compact representations of the tree amplitudes
of six-dimensional maximal super-Yang-Mills theory.
For example, at four and five points the amplitudes are~\cite{DHS}
\begin{eqnarray}
  \mathcal{A}_{4}^{\text{tree}}(1,2,3,4)\;\;\;=
-\frac{i}{s_{12} s_{23}} \, \delta^4\Bigl(\sum_{i=1}^4 q_i^A\Bigr)
\delta^4\Bigl(\sum_{i=1}^4 \tilde{q}_{iB}\Bigr)\,,\nn\\
\mathcal{A}_5^{\text{tree}}(1,2,3,4,5)=\;
\frac{\delta^4(\sum_{i=1}^5 q_i^A) 
  \delta^4(\sum_{i=1}^5\tilde{q}_{iB} )}
  {s_{12}s_{23}s_{34}s_{45}s_{51}}\Bigl\{ q_{1}^A(p_2 p_3 p_4
  p_5)_A^{\;\;B}\tilde{q}_{1B}+\text{cyclic}\nn\\ 
  +\frac{1}{2}\left[q_{1}^A\tilde{\Delta}_{2A} +
  q_{3}^A\tilde{\Delta}_{4A} + (q_3+q_{4})^A\tilde{\Delta}_{5A} +
  (\text{chiral \;conjugate})\right]\Bigr\}\,,
\label{supertrees}
\end{eqnarray}
where $\tilde{\Delta}_{2A} = (p_2 p_3 p_4 p_5-p_2 p_5p_4
p_3)_{A}^{\;\;B}\tilde{q}_{2B}$ and chiral conjugation indicates that one exchanges the chirality of the spinors present in the function. The fermionic delta functions are given by 
\begin{eqnarray}
\delta^4 \Bigl(\sum_{i=1}^n q_i^A\Bigr) \equiv
\frac{1}{4!} \, \epsilon_{BCDE} \, \Bigl( \sum_{i=1}^n q_i^B\Bigr) 
\Bigl(\sum_{i=1}^n q_i^C\Bigr) \Bigl(\sum_{i=1}^n q_i^D\Bigr) 
\Bigl(\sum_{i=1}^n q_i^E\Bigr) \,,
\end{eqnarray}
and a similar expression for the chiral conjugate. 

\subsection{Super sewing rules\label{sumover}}

We now illustrate some key features in the sewing of tree amplitudes
utilizing the $N=(1,1)$ on-shell superspace, either for BCFW
recursion or for unitarity cuts. In general, $n$-point tree
amplitudes take the schematic form,
\eqa 
\mathcal{A}_n&\sim &\left[
\delta\Bigl(\sum q^A\Bigr)\delta \Bigl(\sum\tilde{q}_A\Bigr)\right]^4
q^{n-4}\tilde{q}^{n-4}\quad{\rm
for}\;n>3\,, \nonumber \\ 
\mathcal{A}_3&\sim&\left[\delta\Bigl(\sum
q^A\Bigr)\delta\Bigl(\sum\tilde{q}_A\Bigr)\right]^2
\Bigl(\sum w\cdot\eta\Bigr) \Bigl(\sum \tilde{w}\cdot\tilde{\eta}\Bigr)\,,
\label{face}
\eqae 
where the variables $w$ are additional variables introduced due to the
special kinematics of a three-point amplitude~\cite{CheungOConnell}.
We will not need the precise form here, as we only wish to explain the
structure.  When sewing amplitudes together, we must sum over all
physical states that are allowed to propagate across the cuts. As in
four dimensions, this sum is taken care of by integrating out the
$\eta,\tilde{\eta}$ of the lines being cut.

The presence of super-momentum delta functions again indicates that
one can algebraically solve the delta-function constraints and
substitute the solutions into the remaining functions. This can be
further simplified by combining the delta functions such that a factor
$\delta^4(\sum_{\cal E} q)\delta^4(\sum_{\cal E}\tilde{q})$,
representing the overall supermomentum conservation over the external
legs ${\cal E}$, is extracted outside of the integral. Each of the
remaining delta function pairs is of degree eight, and
hence can be used to localize two pairs of $\eta_l,
\tilde{\eta}_l$s.{\footnote{Recall that
    each $\eta,\tilde{\eta}$ has two components, this gives a total of
    eight components to be localized by the delta function.}}
To implement the localization, it is convenient to
rewrite the delta functions as
\begin{equation}
\delta^4(q_i^A+q_j^A+Q^A)=s_{ij}\delta^{2}\left(\eta_{ia}+\frac{\langle
i_a|\s{p}_jQ}{s_{ij}}\right)\delta^{2}\left(\eta_{ja}+\frac{\langle
j_a|\s{p}_iQ}{s_{ij}}\right),
\label{cutrules}
\end{equation}
where $\eta_i$ and $\eta_j$ are the $\eta$s
that the delta functions localize.

We thus have a general procedure to evaluate the supersums in 
six-dimensional unitarity cuts~\cite{SixD}:
\begin{itemize}
  \item For a cut diagram with $m$ tree amplitudes, choose $m-1$ of
    them. These will be the amplitudes whose supermomentum delta
    function will be used to localize the Grassmann integrals.
    The remaining delta function
    will become the overall delta function that is pulled outside of
    the integral.
  \item Choose $2(m-1)$ independent loop supermomenta for the delta
    functions to localize. (One should choose
    the loop momenta so that the substitution of the solution does not
    imply iterative loops, for example, one delta function enforcing
    $\eta_{l_1}\rightarrow \eta_{l_3}$ while another enforcing
    $\eta_{l_3}\rightarrow \eta_{l_1}$.)
  \item Using eq.~(\ref{cutrules}) we can then easily carry
    out all Grassmann integrations and substitute the solution into the
    remaining functions.
\end{itemize}  
For cuts that contain three-point subamplitudes, the supersum is more
involved due to the introduction of additional variables because of
the special kinematics.  However, we can always choose a spanning set
of cuts with no three-point tree amplitudes or alternatively group
together any three point amplitudes appearing in the cuts with other
tree amplitudes, so as to avoid this technicality.  We refer to
ref.~\cite{SixD} for details on dealing with this case.

\subsection{One-loop four-point example}
We now illustrate the evaluation of supersums by computing the
one-loop four-point maximal super-Yang-Mills in six dimensions. 
Consider again the two-particle cuts in \fig{TwoParticleCutFigure}(a).
Using Grassmann integration to sum the internal states we have
\begin{eqnarray}
&&\sum_{s_1,s_2}\mathcal{A}_{4}(-l_1^{s_2},1,2,l_3^{s_1})\mathcal{A}_{4}(-l_3^{s_1},3,4,l_1^{s_2})\\
&&\nonumber \hskip .3 cm
 =-\int d^2\eta_{l_1}d^2\eta_{l_3}d^2\tilde{\eta}_{l_1}d^2\tilde{\eta}_{k_3}\frac{\delta^4(\sum_R q^A)\delta^4(\sum_R\tilde{q}_A)}{(p-k_1)^2s}\frac{\delta^4(\sum_L q^B)\delta^4(\sum_L\tilde{q}_B)}{s(p+k_4)^2}\,.
\end{eqnarray}
where the $\sum_L$ and $\sum_R$ signify sums over the 
legs on the left and right sides of the cut 
in \fig{TwoParticleCutFigure}(a).
Pulling out the overall supermomentum delta function we arrive at
\begin{eqnarray}
\hskip - 1.5 cm 
\frac{\delta^4(\sum_{i=1}^4 q_i^A)
\delta^4(\sum_{i=1}^4\tilde{q}_{iA})}{(p-k_1)^2s}
\frac{(l_1\cdot l_3)^2}{s(p+k_4)^2}
=-ist \, \mathcal{A}_{{\rm tree}}(1,2,3,4)\frac{1}{(p-k_1)^2(p+k_4)^2} \,,
\label{D6CutResult}
\end{eqnarray} 
where we used $l_3-l_1=k_1+k_2$, and the explicit form of the
four-point superamplitude in \eqn{supertrees} to obtain the final
line.  This answer matches the naive continuation of the
four-dimensional result (\ref{D4CutResult}) to six dimensions.  Thus,
when extending the four-dimensional expressions either to dimensional
regularization or to massively regularized expressions, all terms are
properly captured. The situation is not so simple for
cases with fewer supersymmetries or with larger numbers of loops and
legs.  In these cases the six-dimensional formalism can ensure that 
no contributions are dropped in the regularization or when extending
the results to higher dimensions.

\section{Carrying properties from trees to loops}

We now explain how the unitarity method helps us to carry over novel
tree-level properties to loop level. In particular, symmetries of
tree-level amplitudes can be carried over to loop level if one can
show that all unitarity cuts, combined with the cut propagators,
transform in a universal covariant fashion.  In doing so, one also needs
to track the effect of regularization.
 In this section we will use dual conformal symmetry
as an example showing how the unitarity method
can be used to carry over tree properties to loop level
and to illustrate regularization issues.

A more complex situation arises when the symmetry or property
is directly affected by the cut conditions or when the property is not
global.  This makes it more difficult to prove that the property
carries over to loop level.  Nevertheless, the unitarity method offers
a powerful tool for studying the property at loop level and for
formulating conjectures.  We shall illustrate this situation using the
recently discovered duality between color and
kinematics~\cite{BCJ,BCJLoop,HenrikJJReview}.

\subsection{Dual conformal symmetry}

As discussed earlier, dual conformal symmetry is a conformal symmetry
in the dual $x$-space, where $x$ is defined in terms of momenta, as
given in \eqn{xdefine}. This can be extended to a dual superconformal
symmetry by introducing the fermionic
coordinates~\cite{RecentOnShellSuperSpace},
\begin{equation}
\theta^{\alpha a}_{i}-\theta^{\alpha a}_{i+1}=\lambda_i^\alpha\eta^a_{i}\,.
\label{thetadefine}
\end{equation}  
Once the superamplitudes are written in these dual super coordinates,
the dual conformal property can be straightforwardly established through the
relationship between the conformal boost generator and the translation
generator,
\begin{equation}
K_{\mu}=IP^\mu I\,,
\end{equation}
where $I$ is the inversion operator. The inversion operator acts on
$x$ and $\theta$ as
\begin{equation}
I\left[x_{i\alpha\dot\beta}\right]=\frac{x_{i\beta\dot\alpha}}{x_i^2}\,,
\hskip 1 cm 
I\left[\theta_i^{\alpha a}\right]=
     \frac{x_i^{\dot\alpha \beta}}{x_i^2}\theta_i^a\,_{\beta}\,.
\end{equation}
The inversion properties of the original on-shell variables,
$(\lambda^\alpha,\tilde{\lambda}^{\dot\alpha},\eta^a)$, can be
determined by demanding that inversion preserves the constraints
(\ref{xdefine}) and
(\ref{thetadefine})~\cite{RecentOnShellSuperSpace}. For later use we
note that $\lambda_i$ inverts as\footnote{Our inversion rules
  correspond to choosing $\kappa_i$ of eq.(4.1) of
  ref.~\cite{DualConformalTree} to be $\sqrt{x_i^2x^2_{i+1}}$.}
\begin{equation}
I\left[\lambda^\alpha_{i}\right]=\frac{x_{i}^{\dot\alpha\beta}\lambda_{i\beta}}{\sqrt{x_i^2 x^2_{i+1}}}=\frac{x_{i+1}^{\dot\alpha\beta}\lambda_{i\beta}}{\sqrt{x_i^2 x^2_{i+1}}}\,.
\end{equation}
Since our focus will be on loop amplitudes, where
loop momenta do not 
follow cyclic orderings and we cannot alway label the dual
variables consecutively,
we use the notation
\begin{equation}
x_i-x_j=p_{\{ij\}},\;\;\theta_i-\theta_j=q_{\{ij\}} \,.
\end{equation}  

Ref.~\cite{RecentOnShellSuperSpace} proved, using super-BCFW recursion
relations~\cite{DualConformalTree,AHCKGravity}, that the $n$-point
tree-level amplitude of $\NeqFour$ super-Yang-Mills transforms
covariantly under dual conformal inversion as
\begin{equation}
I\left[\mathcal{A}^{\tree}_n\right]=\Biggl(\prod^n_{i=1}x_i^2\Biggr)
 \mathcal{A}^{\tree}_n\,.
\label{LoopInversion}
\end{equation}
Here we show that it is straightforward to extend this result
to a planar loop-level amplitude statement, via the unitarity
method~\cite{DualConformalTree, SixD, DennenDual},
\begin{equation}
I\left[\mathcal{A}^{(L)}_n\right]=\Biggl(\prod^n_{i=1}x_i^2\Biggr)
 \mathcal{A}^{(L)}_n\,.
\end{equation}
At this point $\mathcal{A}^{(L)}_n$ is understood as the $L$-loop amplitude
prior to integration, defined without a regulator.  Below we describe
the effect of the regulator.  Our discussion will follow the analogous
$D=6$ discussion of ref.~\cite{DennenDual}.

To study the inversion properties in the unitarity cuts, it is more
convenient to separate the (super)momentum delta functions from the
rest of the tree amplitude. For amplitudes with four or more 
external legs, we have
\begin{equation}
\mathcal{A}^{\tree}_n=\delta^4(P)\, \delta^8(Q)f_n,\;\;{\rm with }\;\;
I\left[f_n\right]=\Biggl(\prod^n_{i=1}x_i^2\Biggr)f_n\,.
\end{equation}

We now convert the supersum, which are integrations over $\eta$s, into
integrations over $\theta$. We first note that the total number
of dual points for a loop amplitude is $F=n+L$, where $L$ is the loop
order and $n$ the number of external legs.  The supercut is given by
\begin{eqnarray}
  \mathcal{A}_n^{L}\Bigr|_{\hbox{\footnotesize{cut}}} &=& \int 
    \prod_{\{ij\}} d^4\eta_{\{ij\}} \times
    \mathcal{A}^{\tree}_{(1)}\mathcal{A}^{\tree}_{(2)}\mathcal{A}^{\tree}_{(3)}\ldots\mathcal{A}^{\tree}_{(m)} \nn\\
  &=& \delta^4(P) \int 
   \prod_{\{ij\}} d^4\eta_{\{ij\}} \times
    \prod_{\alpha} \delta^8\left(Q_{\alpha}\right) f_{\alpha}\,,
    \label{FirstStage}
\end{eqnarray}
where we assume the cut is part of a spanning set involving only
tree amplitudes with four or more external legs.
The product over $\{ij\}$ only runs over internal cut lines while the
product over $\alpha$ runs over the tree subamplitudes. Next we use
the identity,
\begin{equation}
 \prod_{\alpha} \delta^8\left(Q_{\alpha}\right)=\int \left[\prod_k d^8\theta_k\right]\left[ \prod_{\{rs\}}\delta^8\left(\theta_r-\theta_s-\lambda_{\{rs\}}\eta_{\{rs\}}\right)\right]\,,
\label{identity}
\end{equation}
where the product over $\{rs\}$ runs over all lines. Since the
translation from supermomentum to dual $\theta$ coordinates has an
overall shift symmetry, the integration measure $d^8\theta_k$ is
understood to include only $F-1$ of them.  To see the equality
(\ref{FirstStage}), one notes that on the LHS the delta functions can
be used to localize the $8(F-1)$ integrals. Denoting the total number
of lines, internal or external, as $P$, there are then $8(P-F+1)$
delta functions left. For planar diagrams, $P=T+F-1$, where $T$ is the
number of tree amplitudes in the cut. Thus, one is left with the
supermomentum delta functions for each tree amplitude, as claimed.

Replacing $\delta^8(Q_{\alpha})$ in \eqn{FirstStage} by
\eqn{identity}, we can now integrate the $\eta$s. Using the super
delta functions to replace the $\eta$s in $f_\alpha$ with $\theta$s,
the remaining $\eta$ dependence then comes solely from the delta
functions. The $\eta$ integral then simply gives
\begin{eqnarray}
  \mathcal{A}_n^{L}\Bigr|_{\hbox{\footnotesize{cut}}} &=& \delta^4(P) \int
    \prod_k d^8\theta_k\times 
    \prod_\alpha f_\alpha \times
    \prod_{\{ij\}} \delta^4(\theta_{ij}\lambda_{\{ij\}})\nn\\
 &&\times\prod_{\{rs\}}\delta^8(\theta_r-\theta_s-\lambda_{\{rs\}}\eta_{\{rs\}})
 \,,
\end{eqnarray}
where now $\{rs\}$ only runs over the external lines. Pulling out an
overall supermomentum delta function, the remaining $8(n-1)$ delta
functions in the product over $\{rs\}$ can be used to localize part of
the $\theta_k$ integration, leaving only the internal dual points. We
are finally left with
\begin{eqnarray}
  \mathcal{A}_n^{L}\Bigr|_{\hbox{\footnotesize{cut}}}=\delta^4(P)\delta^8(Q) \int \biggl(\prod_{l_k} d^8 \theta_{l_k} \biggr) 
  \prod_{\{ij\}} \delta^4(\theta_{ij}\lambda_{\{ij\}})
  \prod_\alpha f_\alpha \,, 
\label{cutintegrand}
\end{eqnarray}
where the product over $l_k$ now runs only over the internal dual
point labels. To calculate the inversion weight of a given cut, we
follow the weight of each contributing piece:
\begin{itemize}
  \item For every internal loop label $l_k$, the $\theta_{l_k}$ measure
    contributes a factor $(x_{l_k}^2)^4$.
  \item Each internal leg $\{ij\}$ contributes $(x_i^2x_j^2)^{-2}$
    which comes from
    $\delta^4\left(\theta_{ij}\lambda_{\{ij\}}\right)$.
  \item Each tree-level subamplitude $f_\alpha$ contributes $\prod_i
    x_i^2$, where $i$ runs over all tree dual labels.
\end{itemize}
After restoring the cut propagators, which invert as
\begin{equation}
I\left[\frac{1}{p^2_{\{ij\}}}\right]=I\left[\frac{1}{x^2_{ij}}\right]=\frac{x_i^2x_j^2}{x^2_{ij}}\,,
\end{equation}
and combining with the other inversion factors, we see that each
planar cut inverts with the overall prefactor,
\begin{equation}
  \Biggl(\prod_{i=1}^n x_i^2\Biggr)\Biggl(\prod_{k=1}^{L} (x_{l_k}^2)^4\Biggr)\,,
\end{equation}
where $x_{l_k}$ are internal dual points. One can extend this
result to cuts involving three-point subamplitudes, though as noted
above this is unnecessary because spanning sets of cuts without
these exist.

If we ignore regularization issues and include the four-dimensional
loop integration measure, the measure will provide an extra inversion
weight of $\prod_k (x_{l_k}^2)^{-4}$, which exactly cancels the extra
weight of internal dual points coming from the integrand.  In any case, the
key result is that all cuts have the same inversion weight. Because
there is a uniform inversion for all cuts, we conclude that the
$L$-loop amplitude prior to integration must invert as given in
\eqn{LoopInversion}.

As noted earlier, for the discussion to hold after integration we need
a regulator, given the presence of infrared divergences.
Fortunately, for $\NeqFour$ super-Yang-Mills, there are natural and
easy to implement regulators.  One option is to use dimensional
regularization and the other a massive regulator~\cite{HiggsReg}.  In
either case, the simplest way to study the dual conformal
properties is to use the six-dimensional helicity and superspace formalism. 
With the six-dimensional language 
we then have simple way to prove the dual conformal properties in the
regularized $D=4$ theory to all loop orders, using generalized unitarity

As shown in ref.~\cite{DennenDual}, following similar steps as the
four-dimensional case outlined above, the six-dimensional integrand
for the $L$-loop amplitude inverts in the same fashion as the
four-dimensional integrand.  To do so we simply need to extend the
dual conformal transformations to act also on the extra-dimensional
components~\cite{SixD,OConnelDualConf} (or equivalently
masses~\cite{HiggsReg}).  Inserting back the loop momentum integral
and the overall momentum and supermomentum delta functions, the
inversion weight differs from \eqn{LoopInversion} by extra powers of
the $x_i$. These are due to solely the extra components in the loop
integration measure and the momentum delta functions.

On the other hand, if we treat the extra dimensional components as
masses, then the loop-momentum integration measure factors and
momentum delta functions are identical to the four-dimensional case.
From the four-dimensional point of view, the momentum of the extra
dimensions is to be identified as regulator masses, and hence should
not be integrated over.  Thus, as shown in ref.~\cite{DennenDual}, the
massively regulated amplitude inverts as in \eqn{LoopInversion}, and
dual conformal symmetry holds in massively regulated planar $\NeqFour$
amplitudes to all loop orders, as expected~\cite{HiggsReg}.  This may
contrasted with dimensional regularization, where the symmetry is
anomalous because the inversion weights of the internal dual points do
not properly cancel.

\subsection{A duality between color and kinematics}

As a second more intricate example,
 we consider the question of carrying over a newly
discovered duality between color and kinematics from tree level 
to loop level.  (For
further details on this duality see the chapter in this review
from Carrasco and Johansson~\cite{HenrikJJReview}.) 

To understand the proposed duality, we first rearrange an $L$-loop
amplitude with all particles in the adjoint representation into
the form,
\begin{eqnarray}
  (-i)^L {\cal A}^\Loop_n  &=& 
\sum_{j}{\int{\prod_{l = 1}^L {d^D p_l \over (2 \pi)^D}
  \frac{1}{S_j}  
 \frac {n_j c_j}{\prod_{\alpha_j}{p^2_{\alpha_j}}}}}\,, \label{LoopBCJ}
\end{eqnarray}
where the sum runs over the set of $n$-point $L$-loop diagrams with
only cubic vertices. These include distinct permutations of external
legs, and the $S_j$ are the symmetry factors of each diagram.  The
form (\ref{LoopBCJ}) can be obtained straightforwardly, for example,
from Feynman diagrams, by representing all contact terms as inverse
propagators in the kinematic numerators that cancel propagators.  We
suppress factors of the coupling constant for convenience.  The
product in the denominator runs over all propagators of each cubic
diagram.  The $c_i$ are the color factors obtained by dressing every
three vertex with an $\f^{abc} = i \sqrt{2} f^{abc}$ structure
constant, and the $n_i$ are kinematic numerator factors depending on
momenta, polarizations and spinors. For supersymmetric amplitudes
expressed in superspace, there will also be Grassmann parameters in the
numerators.

\begin{figure}[tb]
\begin{center}
\includegraphics[clip,scale=0.75]{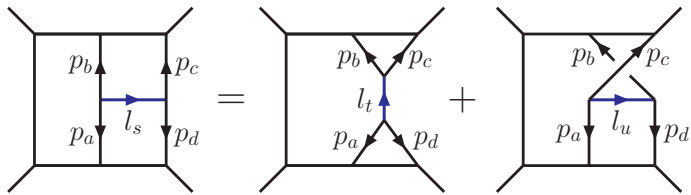}
\end{center}
\vskip -.1 cm 
\caption[a]{\small A numerator duality relation at three loops.  The
relation is either for the color factors or for the diagram
numerators.  }
\label{loopJacobFigure}
\end{figure}

According to the color-kinematics duality proposal of refs.~\cite{BCJ,
  BCJLoop}, arrangements of the diagrammatic numerators in
\eqn{LoopBCJ} exist such that they satisfy equations in one-to-one
correspondence with the color Jacobi identities.  That is, for every
color Jacobi identity we have a relation between kinematic numerators,
\begin{equation}
c_i = c_j-c_k \;  \Rightarrow \;  n_i = n_j-n_k \,.
\label{BCJDuality}
\end{equation}
For example, as illustrated in \fig{loopJacobFigure}, the
numerators of the three displayed diagrams satisfy a similar 
equation as satisfied by the color factors of the diagram.

Perhaps more remarkable than the duality itself, is a related
conjecture that once the gauge-theory amplitudes are arranged into a
form satisfying the duality (\ref{BCJDuality}), corresponding gravity
amplitudes can be obtained simply by taking a double copy of
gauge-theory numerator factors~\cite{BCJ,BCJLoop},
\begin{eqnarray}
 (-i)^{L+1}  {\cal M}^\Loop_n  &=& 
 \sum_{j} {\int{ \prod_{l = 1}^L {d^D p_l \over (2 \pi)^D}
 \frac{1}{S_j}
   \frac{n_j \n_j}{\prod_{\alpha_j}{p^2_{\alpha_j}}}}} \,,
 \label{LoopBCJGrav}
\end{eqnarray}
where the $\n_i$ represent numerator factors of a second gauge theory
amplitude, the sum runs over the same set of diagrams as in
\eqn{LoopBCJ}.  We suppressed the gravitational coupling constant in
this expression.  This is expected to hold in a large class of
gravity theories, including theories that are the low-energy limits of
string theories.  It should also hold in pure gravity, but in this
case extra projectors would be required to remove the unwanted states
arising in the direct product of two pure Yang-Mills theories.  At
tree level ($L=0$), this double-copy property is closely related to
the KLT relations between gravity and gauge
theory~\cite{KLT}.  

For the tree case, through use of BCFW recursion, the double-copy
formula (\ref{LoopBCJGrav}) has been proven for pure gravity and for
$\NeqEight$ supergravity tree amplitudes in $D=4$, assuming the
duality~(\ref{BCJDuality}) holds in the corresponding gauge
theories~\cite{Square}.  At tree level a consequence of this duality
is non-trivial relations between the color-ordered partial amplitudes
of gauge theory~\cite{BCJ,Bjerrum1,Feng}.  The duality has also been
studied in string theory~\cite{Bjerrum2,Tye}.  This duality has also
been discussed from the Lagrangian vantage point in
ref.~\cite{Square}. Recently, an alternative trace-based
representation of the color-kinematics duality was given
in~\cite{Trace}.

The non-trivial part of the all-loop conjecture (\ref{BCJDuality}) is
that there is sufficient freedom to arrange gauge-theory multiloop
amplitudes in a way that satisfies the color-kinematics duality
(\ref{BCJDuality}).  To analyze whether the conjecture is plausible,
we can use the unitarity method.

The present situation is more complicated than the previous example of
dual conformal symmetry. In the dual conformal case, when intermediate
momenta are placed on shell, the properties are essentially unchanged.
This allows the dual conformal properties to easily be carried to loop
level, since they hold globally for all cuts.  In the present case,
when a line is cut, we lose the Jacobi-like relation involving that
line, altering the system of equations enforcing the duality.  In
addition, the remaining numerator relations become simpler because of
the extra on-shell cut conditions.  The duality (\ref{BCJDuality}) is
implemented as a collection of identities which act locally on diagrams,
but which have global consistency implications. The cuts do not need
to satisfy the complete set of duality relations satisfied by the loop
amplitude, but only a subset.  This makes proving that the
color-kinematics duality holds for loop amplitudes more difficult than
proving that dual conformal covariance holds in the planar case.  Indeed,
we still do not have a proof that the color-kinematics duality holds
to all loop orders.  Whether it does depends on the balance between
the number of constraints and the freedom to rearrange the amplitudes
as the loop order increases.  Nevertheless, the fact that the duality
(\ref{BCJDuality}) holds for {\it all} cuts decomposing loop amplitudes
to sums of products of tree amplitudes, assuming it holds at tree level, is a
key motivation for extending it to all loop orders~\cite{BCJ,BCJLoop}.

Of course, given that we do not yet have a proof that the
color-kinematics duality holds, it is important to check it in explicit
examples.  The unitarity method has been used to provide such
examples.  The one- and two-loop four-point amplitudes of $\NeqFour$
super-Yang-Mills theory and $\NeqEight$ supergravity, as obtained in
ref.~\cite{BDDPR}, are easily shown to
satisfy the conjecture (\ref{BCJDuality}).  Another example is the
two-loop four-point identical-helicity amplitude of pure Yang-Mills
theory~\cite{TwoLoopAllPlus}, which also has been shown to satisfy the
duality~\cite{BCJ,BCJLoop}.  Finally, a rather nontrivial example is the
three-loop four-point amplitude of $\NeqFour$ super-Yang-Mills
theory~\cite{BCJLoop}. This example is discussed further in another
chapter of this review~\cite{HenrikJJReview}, including a discussion
of how the color-kinematics duality can be exploited to simplify
calculations of the generalized cuts.

\section{Concluding Comments}

In summary, generalized unitarity is an established tool for carrying
out loop calculations in gauge and gravity theories, having led to a
variety of new nontrivial results.  Here we summarized four-dimensional
and six-dimensional versions, making use
of helicity and on-shell superspaces. The six-dimensional 
version offers a means of regularizing amplitudes, while retaining
key advantages of the four-dimensional version.
The unitarity method also offers a straightforward means for
carrying any identified tree-level property to loop level. As particular
examples, we discussed dual conformal symmetry as well as a new
duality between color and kinematics.  We expect that in the
coming years, generalized unitarity will continue to play a leading 
role in new calculations and for uncovering exciting new properties of
scattering amplitudes.

\ack We are grateful to John Joseph Carrasco, Scott Davies, Tristan Dennen,
Lance Dixon, Fernando Febres Cordero,
Darren Forde, David Kosower, Harald Ita, Henrik Johansson and
Radu Roiban for related collaborations and for many stimulating
discussions on the topics described here.

\end{document}